\documentclass[11pt,a4]{article}
\pdfoutput=1
\usepackage{jheppub}
\usepackage{epsfig}
\usepackage{amsfonts}
\usepackage{amsmath}
\usepackage{graphicx}
\usepackage{color}
\usepackage{amssymb,amsmath,hyperref,slashed,color,graphicx,bm}
\usepackage[T1]{fontenc}

\hypersetup{colorlinks,citecolor= nicegreen,linkcolor= nicered}
\definecolor{nicered}{rgb}{0.7,0.1,0.1}
\definecolor{nicegreen}{rgb}{0.1,0.5,0.1}

\newcommand{\be}{\begin{equation}}
\newcommand{\ee}{\end{equation}}
\newcommand{\bea}{\begin{eqnarray}}
\newcommand{\eea}{\end{eqnarray}}

\newcommand{\bc}{\begin{center}}
\newcommand{\ec}{\end{center}}

\newcommand{\half}{\frac{1}{2}}

\usepackage{graphicx,amssymb,amsmath,amsfonts}

\title{Extending the scope of holographic mutual information and chaotic behavior}

\preprint{TAUP-3006/16}

\author{Nilanjan Sircar, Jacob Sonnenschein and Walter Tangarife}

\affiliation{The Raymond and Beverly Sackler School of Physics and Astronomy,\\
	Tel Aviv University, Ramat Aviv 69978, Israel}
\emailAdd{nilanjan.tifr@gmail.com}
\emailAdd{cobi@post.tau.ac.il}
\emailAdd{waltert@post.tau.ac.il}

\abstract{
We extend the use of holography to investigate the scrambling properties of various  physical systems. Specifically, we consider: (i) non-conformal backgrounds of black $Dp$ branes, (ii) asymptotically Lifshitz black holes, and (iii) black $AdS$ solutions of  Gauss-Bonnet gravity. We use the disruption of  the entanglement entropy as a probe of the chaotic features of such systems. Our analysis shows that these theories share the same type of behavior as conformal theories as they undergo chaos; however, in the case of Gauss-Bonnet gravity, we find a stark difference in the evolution of  the mutual information for negative Gauss-Bonnet coupling. This may signal an inconsistency of the latter.
}


\begin{document}
\maketitle
\flushbottom

\section{Introduction and summary}

In recent years, there has been an increasing convergence of interests among the quantum information, high energy theory and condensed matter communities. A common ground of this interdisciplinary interaction is the use of entanglement entropy (EE) to characterize quantum systems. In the gauge/gravity correspondence, the entanglement entropy of a boundary region $A$ is determined by the area of the minimized codimension-2 bulk surface that coincides with the entangling surface at the boundary \cite{Ryu:2006bv,Ryu:2006ef,Hubeny:2007xt}. An extra ingredient is the association of the properties of a black hole in the bulk with a thermal state in the boundary theory, providing a powerful tool to answer questions regarding the thermalization of strongly coupled theories.

Equilibration in thermal systems has been studied in holography in a variety of settings via EE. Some of the most relevant approaches include calculating quasi-normal modes of small perturbations in the bulk geometry to determine the relaxation rate of the perturbation (near equilibrium) \cite{Horowitz:1999jd} and following the formation of a black hole in the bulk  (far from equilibrium) \cite{Balasubramanian:2011ur}. The latter approach has been used to further understand the time evolution of entanglement entropy and mutual information as a system thermalizes \cite{Balasubramanian:2011ur,Allais:2011ys,Callan:2012ip,Galante:2012pv,Caceres:2012em,Caceres:2013dma,Liu:2013iza,Liu:2013qca,Kundu:2016cgh}.

Black holes are conjectured to be {\it fast scramblers}, meaning that they thermalize faster than other systems in nature \cite{Hayden:2007cs, Sekino:2008he}. Specifically, the so-called scrambling time scales logarithmically with the black hole entropy,
\begin{equation}
t_*\,\sim \,\frac{\beta}{2\pi} \,\log\,S_{BH}.
\end{equation}
This is the time scale at which all correlations in a system have been destroyed after introducing a perturbation. The scrambling time plays a crucial role in the evolution entanglement entropy. In \cite{Shenker:2013pqa}, this was illustrated by looking at the time evolution of the two-dimensional thermo-field double state (TFD)
\begin{equation}
| \Psi \rangle \,\equiv \, \sum_i\,e^{-\beta E_i/2}  | i \rangle_L \otimes  | i \rangle_R,
\end{equation}
where $  | i \rangle_L$ and $ | i \rangle_R$ are identical states of two quantum theories. The holographic dual of such a pure state is given by the eternal $AdS$ black hole in $2+1$ dimensions \cite{Maldacena:2001kr}. A perturbation to this state is introduced by adding a shock wave, with energy $E$ at a boundary time $t_w$, to the two-sided AdS black hole geometry. One can then characterize the dependence between the two boundaries by computing mutual information~\footnote{The mutual information between subsystems in two different CFTs
of the thermofield double was introduced and studied in \cite{Morrison:2012iz} and coined as {\it thermo-mutual information}.},
\begin{equation}
I(A,B)\,\equiv \, S_A\,+\,S_B\,-\,S_{A\cup B}, \label{Eq:MI_def}
\end{equation} where $A$ and $B$ are space-like regions in each boundary. In the specific case of a $2+1$ dimensional bulk, $S_A$ and $S_B$ are given by the length of the geodesic that extends between the boundary of the intervals $A$ and $B$, respectively. The geodesic used to compute $S_{A\cup B}$ is one traversing across the black hole horizon, joining the two boundaries. For large intervals ($\sinh(\frac{\pi \phi \ell}{\beta})>1$, $\phi<\pi$), the resultant mutual information is given by
\begin{equation}
I(A,B)\,=\,\frac{\ell}{G}\left(\log\, \sinh\,\frac{\pi\phi \ell}{\beta}\,-\,\log \left(1+\frac{E\,\beta}{4S}e^{2\pi\,t_w/\beta}  \right)   \right), \label{Eq:MISS}
\end{equation}
where $\phi$ is the angular size of the $A$ and $B$,  $\ell$ is the $AdS$ radius, and $S$ is the black hole entropy. From this expression, it can be seen $I(A,B)$ vanishes when (for $\beta \gg \ell$),
\begin{equation}
t_w\,=\,t_*\,=\,\frac{\phi\,\ell}{2}\,+\,\frac{\beta}{2\pi}\log \,\frac{2S}{\beta E}.
\end{equation} This disruption of the entanglement, which corresponds to the saturation of mutual information, is evidence of the state's high sensitivity to variations of initial conditions, a phenomenon known as {\it the butterfly effect}. Because mutual information acts as an upper bound on the correlations between the subsystems, the scrambling time defined in this fashion sets the time at which any sort of dependence between the $A$ and $B$ has been erased \cite{Swinney,Caputa:2015waa,Hosur:2015ylk}. This analysis has been generalized to $AdS_d$ in \cite{Leichenauer:2014nxa}, multiple shock waves \cite{Shenker:2013yza}, and localized shock waves \cite{Roberts:2014isa, Caputa:2015waa}. Likewise, the stringy corrections of the scrambling time were presented in \cite{Shenker:2014cwa}.

In a follow up work, a bound on how fast a thermal quantum field theory (QFT) can develop chaos was conjectured in \cite{Maldacena:2015waa}. Based on the results from Einstein gravity calculations \cite{Shenker:2013pqa}, a bound on the largest Lyapunov exponent was proposed $\lambda_L\,\leq \frac{2\pi\ k_B T}{\hbar}$  \footnote{The Lyapunov exponent was associated to the damping coefficient of the thermal gluon plasma using holography in \cite{KalyanaRama:1999zj}}.

In this article, we extend the scope of the use of holography to investigate  the chaotic behavior, and in particular the scrambling time, of various systems which have not yet been explored: (i) The non-conformal backgrounds of black Dp branes\cite{Itzhaki:1998dd}. (ii) Black holes of  asymptotically Lifshitz non-relativistic backgrounds\cite{Balasubramanian:2009rx} (iii) Conformal backgrounds which are  solutions of the Gauss-Bonnet gravity \cite{Cai:2001dz,Myers:2010ru,deBoer:2011wk,Hung:2011xb,Myers:2012ed}.

Our analysis is based on the idea, suggested in \cite{Shenker:2013pqa}, of using the disruption of  the entanglement entropy  as the probe of chaos. We apply the procedure and techniques  presented in \cite{Leichenauer:2014nxa} on the  three classes of background mentioned above.  The procedure includes the following steps: (i) We compute the entanglement entropy $S_A$ and $S_B$ associated with a strip of width $L$. This includes the renormalization of the EE's by subtracting the divergent parts of them. (ii) We compute the mutual information $I(A,B)$. (iii) We introduce a null perturbation in the form of a shock wave and determine the corresponding backreacted geometry.
(iv) We re-express  the metric in  Kruskal coordinates $u$ and $v$. The shock wave propagates along a constant $u$ and causes a shift in $v$ denoted by  $\alpha$ (see (\ref{Eq:alpha_gen})). We compute  $\alpha$   and get an approximation of  the scrambling time $t_*$ from the condition of $\alpha \sim 1$. (v) We compute the dependence of $\alpha$ on $r_0$, the  value of the radial coordinate of the tip of the entangling surface, and then obtain the mutual information $I(A_h,B_h)$ associated with the shock wave as a function of $\alpha$; in particular, we obtain numerically $\alpha_*$, where $I$ vanishes. For the cases of the $Dp$-brane and Lifshitz backgrounds, we determine EE using the prescription of \cite{Ryu:2006bv,Hubeny:2007xt} whereas for the GB gravity we adopt the prescription of \cite{deBoer:2011wk} and \cite{Hung:2011xb}.

The outcome of our holographic analysis is the following:
\begin{itemize}
\item
The dual gravitational description of Yang Mills theories in $p+1$ dimensions with 16 supersymmetries in terms of the $Dp$ backgrounds is reliable only in range of values of the effective coupling where the corresponding curvature and string coupling are small \cite{Itzhaki:1998dd}. Our analysis of the mutual information in this range  shows that the behavior for $p\leq 4, \ p\neq 3$  the non-conformal cases  is similar to that of $p=3$ the conformal case. We found that the dependence of $I$ on $\alpha$ becomes steeper upon increasing $p$ for $p<5$. For $p=5$ the situation is just the opposite and the dependence is flatter.  The fact that for $p\geq5$ the dependence is different is consistent with the findings of \cite{Barbon:2008sr} and \cite{Kol:2014nqa}.
\item
For the Lifshitz non-relativistic backgrounds that are characterized by a different scaling of the time and space directions, we found that basically the behavior is like for the relativistically invariant backgrounds. The dependence of $\alpha$ on the width of the interval $L$ is steeper as one increases $z$ and the dependence of $I$ on $\alpha$ is flatter, namely, $I$ vanishes at a larger values of $\alpha$
\item
In the GB gravity,  for positive coupling $\lambda$ which is  in accordance with causality \cite{Brigante:2008gz},    the dependence of  the mutual information in the presence of a  shock wave  on $\alpha$ is smooth and monotonically decreasing,  in a similar manner to that of the usual Einstein gravity case. $\alpha_*$, where the mutual information vanishes increases with the increase of $\lambda$. On the other hand for negative $\lambda$ and in the allowed region  according to \cite{Brigante:2008gz} we encountered an anomalous  ``cusp'' behavior depicted in Figure \ref{Fig:MI-jump}. This follows from the fact that $z_0$ is a double valued function of $\alpha$, namely,  for any given value of $\alpha$ there are two values of $z_0$ and hence two values of the mutual information $I$. Note that whereas around   $z_0$ that hits the singularity  the curvature is large, this is not the case for the point of the ``cusp''.  This type of anomalous behavior was detected also in the computation of the entanglement entropy in \cite{Caceres:2015bkr}. It does not seem likely that it signals a special behavior in the dual boundary field theory but rather that the GB gravity maybe inconsistent. An inconsistency with causality was found in \cite{Camanho:2014apa} though there the causal behavior was found out to for both positive and negative values of $\lambda$ and in our case it is only for the latter.

\end{itemize}
%
The paper is organized as follows: In the next section we describe the determination of the scrambling time in general. Section~\ref{Sec:Non-Conformal} is devoted to the analysis of the chaotic behavior associated with non-conformal backgrounds of the form of Dp branes. In  section~\ref{Sec:DpMetric} the geometry of black Dp branes is reviewed. We then compute the mutual information in these geometries in section~\ref{Sec:MI_NonSW_Dp}. In  section~\ref{Sec:ScramblingTimeDp} we write down  the relation between $\alpha$ and $t_*$ and  determine the latter. The Penrose diagram associated with the minimal surface in the presence of the shock wave  is drawn in  section~\ref{Sec:Ext_Surf_HP_Dp}. We then compute the relation between $\alpha$ and $r_0$. We then determine  the mutual information as a function of $\alpha$ in section~\ref{Sec:MI_SW_Dp}.
Section~\ref{Sec:Nonrelativistic} is devoted to the case of asymptotically Lifshitz non-relativistic background. In section~\ref{Sec:Lif_metric} we review the basic properties of the Lifshitz backgrounds and write down an action of a particular model and its corresponding metric. The mutual information of this system prior to introducing the shock wave is  written down in section~\ref{Sec:MI_NonSW_Lif}. In section~\ref{Sec:ScramblingTime_Lif} we introduce the shock wave and approximate the scrambling time. The description of the extremal surfaces for half plane in TFD is given in section~\ref{Sec:Ext_Surf_HP_Lif} and the calculation of the mutual information in section~\ref{Sec:MI_SW_Lif}.
In  section~\ref{Sec:HD_Theories} we present the analysis of the holographic mutual information in the case of Gauss-Bonnet higher curvature  gravity. In  section~\ref{Sec:HD_metric} we present the action and the corresponding metric solution of the equations of motion.  The mutual information, scrambling time and the  extremal surface are analyzed in subsections  \ref{Sec:MI_NSW_HD}, \ref{Sec:Scrambling-HD} and \ref{Sec:ExtremalSurface_HD} respectively.
In section~\ref{Sec:summary}  we summarize the results of this paper and list several open questions.
We then add in appendix \ref{Appendix:SeriesExp-Dp} a series expansion of EE in Dp-brane backgrounds. In appendix \ref{Appendix:AnalyticExtSurf-D5} we give an analytic
expression for two sided entanglement etropy for half planes in case of $D5$ brane.




\section{Scrambling time in general geometries}\label{Sec:GenSWMetric}

In this section, we summarize the generalities of the scrambling time calculation in an eternal black hole geometry. We follow the analysis of a generic metric presented in the appendix of \cite{Shenker:2013pqa}.

Let us start by considering a $d+1$-dimensional metric of the form,
\begin{eqnarray}
ds^2 &=& -f(r) dt^2+\frac{dr^2}{f(r)} + d\Sigma_{d-1}^2,  \\
d\Sigma_{d-1}^2 &=& g_{ij}(r,x^i) dx^i dx^j.
\end{eqnarray}
We further assume the existence of a non-extremal horizon at $r=r_h$, such that $f(r_h)=0$ and $f'(r_h) \ne 0$.
The inverse Hawking temperature associated with the black hole is given as $\beta =\frac{4 \pi}{f'(r_h)}$.

We then rewrite the metric using the Kruskal coordinates,
\begin{eqnarray}
&& u v =- e^{f'(r_h) r_*(r)} ,~~~~ u/v = e^{-f'(r_h) t},\\
&& ds^2 = -\frac{4 f(r)}{f'(r_h)^2} e^{-f'(r_h) r_*(r)} du dv + d\Sigma_{d-1}^2, \label{Eq:Kruskal-gen}
\end{eqnarray}
where $dr_* = \frac{dr}{f(r)}$ is the so-called tortoise coordinate, which behaves such that $r_*\rightarrow -\infty$ as $r\rightarrow r_h$.

\begin{figure}
\begin{center}
 \includegraphics[scale=.32]{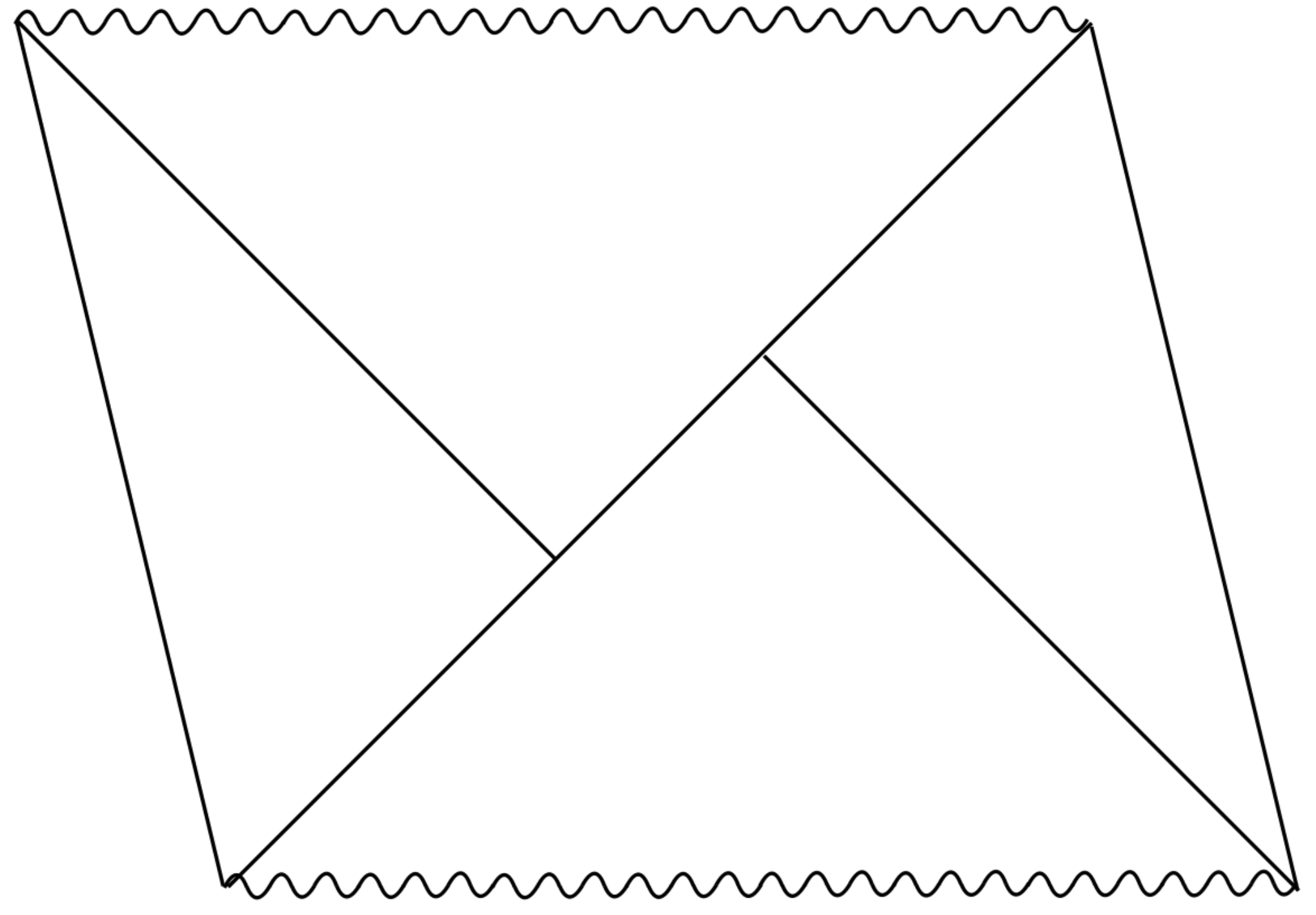}
 \caption{The geometry of an eternal black hole under the perturbation of a shock wave.}\label{Fig:Shockwave}
 \end{center}
\end{figure}

We now add a global null perturbation of
asymptotic energy $E \ll M$ ($M$ is the ADM mass) at time $t_w$ and radius $r=\Lambda$. $\tilde u, \tilde v$ denotes coordinates to the
left of the perturbation and $u, v$ to the right. The shell propagates on a $u={\rm const}$ surface given by,
\be
\tilde u_{w} = e^{\frac{\tilde f'(\tilde r_h)}{2} (\tilde r_*(\Lambda)-t_w)} ~~~ u_{w} = e^{\frac{f'(r_h)}{2} (r_*(\Lambda)-t_w)} .
\ee
We patch the two sides of the geometry along this surface, using the following matching condition,
\be
\tilde u_w \tilde v = - e^{\tilde f'(\tilde r_h) r\tilde{r}_*(\tilde{r})}~~~ u_w v = - e^{f'(r_h) r_*(r)}.
\ee
We expect that at large $t_w$ and order $\mathcal{O} (E)$
\be
\tilde v = v + \alpha .
\ee
The Penrose diagram corresponding to the backreacted geometry is presented in Figure \ref{Fig:Shockwave}.

For small $E$ we can approximate $\tilde u_w =u_w$. Large $t_w$ pushes $r$ to $r_h$, so that we can approximate
$f(r)= f'(r_h) (r-r_h) + \cdots$. Then $e^{f'(r_h) r_*}=(r-r_h) C(r,r_h)$, where $C(r,r_h)$ is smooth and non-zero
at $r= r_h$. Then, to linear order in $E$,
\be
\alpha = \frac{E}{u_w} \frac{d}{dM} \left[ (r_h-r) C(r,r_h)\right]|_{r=r_h}=\frac{E}{u_w} \frac{d r_h}{dM} C(r_h,r_h). \label{Eq:alpha_gen}
\ee
By Bekenstein-Hawking area formula we can relate $r_h$ and  the entropy~\footnote{This result is valid also for higher derivative black holes with planar horizon~\cite{Cai:2001dz}},
\be
S_{BH} = \frac{V_{d-1}(r_h)}{4 G_N} ,
\ee
where, $V_{d-1}(r_h) = \int_{\Sigma} d^{d-1} x \sqrt{g_{\Sigma}}$ is the area of the horizon.
Now we can use the first law,
\be
E = \boxed{\delta M = T \delta S} = T \frac{V_{d-1}'(r_h)}{V_{d-1}(r_h)} S_{BH} \frac{d r_h}{dM} E ,
\ee
which gives,
\be
\frac{d r_h}{dM} = \frac{1}{T}  \frac{V_{d-1}(r_h)}{V_{d-1}'(r_h)} \frac{1}{S_{BH}} ,
\ee
then,
\be
\alpha = \frac{E}{T} C(r_h,r_h) \frac{V_{d-1}(r_h)}{V_{d-1}'(r_h)} \frac{1}{S_{BH}} e^{-\frac{2\pi}{\beta} (r_*(\Lambda)-t_w)} . \label{alpha_genmetric}
\ee
The scrambling time is given by $t_w=t_*$ when $ \alpha \sim 1$~\cite{Shenker:2013pqa}. Inverting that,
\be
t_* = r_*(\Lambda) + \frac{\beta}{2 \pi} \log \left[ \frac{V_{d-1}'(r_h)}{V_{d-1}(r_h)} \frac{T}{E} \frac{1}{C(r_h,r_h)} S_{BH}\right] .
\ee
It is noteworthy that although the coordinate $r_*$ can be shifted by a arbitrary constant, $t_*$ is not changed by it, as
the combination $r_*(\Lambda)-\frac{\beta}{2 \pi} \log C(r_h,r_h)$ is invariant under this shift.
Let us define two dimensionless constant $c_1,~c_2$ depending on the details of the geometry (as illustrated in following sections), 
\be
\frac{V_{d-1}'(r_h)}{V_{d-1}(r_h)} = \frac{c_1}{r_h},\quad {\rm and}\quad C(r_h,r_h)=\frac{c_2}{r_h}.
\ee
For shock wave energy (density in case of non compact horizons) $E \sim T$,
\be
t_* = r_*(\Lambda) + \frac{\beta}{2 \pi} \log S_{BH} +  \frac{\beta}{2 \pi} \log \frac{c_1}{c_2}  .\label{ts_general_heuristic}
\ee
Notice that $S_{BH}$ denotes entropy density in case of non-compact horizons.




\section{The butterfly effect in non-conformal backgrounds}\label{Sec:Non-Conformal}

In this section, we want to study the disruption of mutual information in non-conformal theories in $p+1$ dimensions. These are theories that correspond to black $p-$brane supergravity solutions \cite{Itzhaki:1998dd}. We begin by describing the bulk metric in the Einstein frame and computing the mutual information for the unperturbed geometry. We then proceed to find the scrambling time after adding a shock wave perturbation to the two-sided black hole, and conclude by numerically verifying the disruption of mutual information between the two boundary theories.

\subsection{Black $Dp$-brane geometry}\label{Sec:DpMetric}

We consider the near extremal black $p-$branes whose geometry is given by \cite{Itzhaki:1998dd}
\begin{equation}
ds^2\,=\,-f(r) \,dt^2 \,+\,\frac{dr^2}{f(r)}\,+ \left(\frac{r}{R_p}\right)^{\frac{7-p}{2}} d\vec{y}^2 +R_p^{\frac{7-p}{2}} r^{\frac{3-p}{2}} d\Omega_{8-p}^2,
\end{equation} where
\begin{equation}
f(r)\,\equiv\,
\left(\frac{r}{R_p}\right)^{\frac{7-p}{2}} \left(1- \left(\frac{r_H}{r}\right)^{7-p}\right), \qquad R_p^{7-p} = (4 \pi)^{\frac{5-p}{2}} \Gamma\left(\frac{7-p}{2}\right) g_s l_s^{7-p} N,
\end{equation} and the dilaton field is
\begin{equation}
e^{\phi-\phi_{\infty}}=\left(\frac{R_p}{r}\right)^{\frac{(7-p)(3-p)}{4}}, \quad g_s=e^{\phi_{\infty}}.
\end{equation}
This background is dual to super-Yang-Mills in  $p+1$ dimensions. The non-trivial dilaton corresponds to the running of the coupling constant $g_{eff}^2=g_{YM}^2 N \frac{r^{p-3}}{l_s^{2(p-3)}}$, where  $g_{YM}^2=(2\pi)^{p-2} g_s l_s^{p-3}$. The validity of the supergravity solution requires,
\be
1 \ll g_{eff}^2 \ll N^{\frac{4}{7-p}},
\ee such that both the dilaton and the curvature are small.

For convenience, we will use a coordinate re-parametrization of the Einstein frame in our calculations. The Einstein frame metric is obtained by the following transformation of the metric in $9+1$ dimensions,
\be
g_{\mu \nu}^E=e^{-\frac{\phi-\phi_{\infty}}{2}} g_{\mu \nu}^S.
\ee
Then, the resultant metric is given by
\be
ds_E^2 = -g_1(r) dt^2+\frac{dr^2}{g_2(r)}+\left(\frac{r}{R_p}\right)^{\frac{(7-p)^2}{8}} d\vec{y}_{\parallel}^2
+R_p^2 \left(\frac{r}{R_p}\right)^{\frac{(p-3)^2}{8}} d\Omega_{8-p}^2,
\ee
where,
\bea
g_1 (r) &=& \left(\frac{r}{R_p}\right)^{\frac{(7-p)^2}{8}} \left(1-\left(\frac{r_0}{r}\right)^{7-p} \right), \\
g_2 (r) &=& \left(\frac{r}{R_p}\right)^{\frac{(7-p)(p+1)}{8}} \left(1-\left(\frac{r_0}{r}\right)^{7-p} \right).
\eea

Now, we perform the coordinate redefinition
\be
r = \frac{\rho^n}{n^n R_p^{n-1}} ~,~~ n= \frac{8}{8+(7-p)(3-p)},
\ee
under which the metric in Einstein frame can be written as
\bea
ds_E^2 &=& -f(\rho) dt^2+\frac{d\rho^2}{f(\rho)}+\left(\frac{\rho}{n R_p}\right)^{\frac{n (7-p)^2}{8}} d\vec{y}_{\parallel}^2
+R_p^2 \left(\frac{\rho}{n R_p}\right)^{\frac{n(p-3)^2}{8}} d\Omega_{8-p}^2, \label{metricEF1} \\
f(\rho)&=& \left(\frac{\rho}{n R_p}\right)^{n \frac{(7-p)^2}{8}} \left(1-\left(\frac{\rho_H}{\rho}\right)^{n(7-p)} \right) .
\eea

In this geometry, the temperature and entropy density~\footnote{We have used the relations of $10-$dimensional Newton's constant to $g_s$,$l_s$
:$16 \pi G_N^{(10)}=
(2\pi)^7 g_s^2 l_s^8$.} are given by,
\begin{eqnarray}
 T &&= \frac{(7-p)}{4 \pi R_p} \left(\frac{\rho_H}{n R_p}\right)^{\frac{n}{2}(5-p)} ,  \\
 \frac{S}{\Omega_8 V_p} && = c(p) \lambda^{\frac{p-3}{5-p}} N^2 T^{\frac{9-p}{5-p}} ,\label{Eq:Dp_Entropy}
\end{eqnarray}
where $$c(p)=\frac{2^{\frac{2(11-2p)}{5-p}} \pi^{\frac{19-8p+p^2}{2(p-5)}}\Gamma\left(\frac{7-p}{2}\right)^{\frac{7-p}{5-p}}}{\left(7-p\right)^\frac{9-p}{5-p}},$$
and $\lambda=g_{YM}^2 N$ is the 't Hooft coupling in the dual field theory.

In the following subsections, we will use this metric with the change in notation, for convenience,
\begin{equation}
r \equiv \rho, \quad r_h \equiv \rho_h,\quad {\rm and} \quad \ell \equiv  n\,R_p. \label{Eq:new_rs}
\end{equation}

\subsection{Mutual Information in black $Dp$-brane geometries} \label{Sec:MI_NonSW_Dp}

In this section, we present an overview of the mutual information between the two boundaries of an eternal black hole in the $Dp$-brane background. Specifically, we compute the mutual information (\ref{Eq:MI_def}) between two strips $A$ and $B$ contained in the left and right side of the geometry respectively\footnote{We will also refer to these strips as ``intervals'' in the rest of this article}. To achieve this, we must compute  the EE between each strip and the rest of the system ($S_A$ and $S_B$), and  the EE of their union ($S_{A\cup B}$).

We begin by reviewing the calculation of  the holographic EE of an interval using the background given by (\ref{metricEF1}) with the coordinates defined in (\ref{Eq:new_rs}). Each interval is defined along the coordinates $$y\equiv y^1 \in [-L/2,\,L/2],\,y^{i=2,\cdots, p-1}\in (-\infty,\,\infty).$$ The holographic prescription \cite{Klebanov:2007ws} stipulates that $S_A$ (or $S_B$) is given by the codimension-2 minimal surface, in the bulk, whose boundary is the same as the boundary of $A$ (or $B$). We parameterize this surface by $x=(y(r),\,y^2,\cdots, y^{p-1},\, \theta^1,\cdots, \theta^{8-p})$. Thus, $S_A$ is given by
\begin{equation}
S_A\,=\,\frac{1}{4G^{10}_N} \int d^{8}x \sqrt{{\rm Det\,}G^{(8)}_{\rm ind}},
\end{equation} where $G^{(8)}_{\rm ind}$ is the induced metric on the surface, using (\ref{metricEF1}). This yields
\begin{equation}
S_A\,=\,\frac{\mathcal{V}\ell^{8-p}}{4G^{10}_N\ell^{d-1}} \int dr\,\left[\left(\frac{r}{\ell}\right)^{\frac{n}{16}((7-p)^2(p-1)+(p-3)^2(8-p))}\left(\frac{1}{f}+\left(\frac{r}{\ell}\right)^{\frac{n}{8}(7-p)^2} y'^2 \right)  \right]^{1/2},  \label{Eq:SA_1}
\end{equation} where $'$ denotes the derivative respect to $r$ and $\mathcal{V}$ is the volume in the transverse directions. The absence of $y$ in the Lagrangian allows us to write the conservation equation
\begin{eqnarray}
&&\frac{\left(\frac{r}{\ell}\right)^{\omega/2} y'}{\sqrt{y'^2 +\left(\frac{r}{\ell}\right)^{-2\xi}\frac{1}{f}}}\,=\, \left(\frac{r_{\rm min}}{\ell}\right)^{\omega/2}, \quad 1/y' \big|_{r_{\rm min}}\,=\,0, \label{Eq:EOMDps} \\
&& \omega\,\equiv\, \frac{p\,n}{8}(7-p)^2+ \frac{n}{8}(p-3)^2(8-p),\,\,\xi\,\equiv \, \frac{n}{16}(7-p)^2. \label{Eq:Const-defs}
\end{eqnarray}

Solving for $y'$ gives us the relation between the width $L$ and the turning point $r_{\rm min}$,
\begin{equation}
L\,=\,2\int_{r_{\rm min}}^{\infty} dr \, \left(\frac{r}{\ell}\right)^{-2\xi}  \frac{1}{\sqrt{1-\left(\frac{r_H}{r}\right)^{n(7-p)}} \sqrt{\left( \frac{r}{r_{\rm min}}\right)^{\omega} -1 }}.  \label{Eq:L_Einst}
\end{equation} On the other hand, plugging (\ref{Eq:EOMDps}) into (\ref{Eq:SA_1}) gives us $S_A$ as a function of $r_{\rm min}$,
 \begin{equation}
S_A\,=\,2\frac{\mathcal{V}}{4G_N \ell^{d-1}} \ell^{8-p-\sigma} \int_{r_{\rm min}}^{\infty}dr \frac{r^\sigma}{\sqrt{1-\frac{r_H^s}{r^s}}} \frac{1}{\sqrt{1-\left( \frac{r_{\rm min}}{r}\right)^{\omega}}}, \label{Eq:area_Einst}
\end{equation}
where \begin{eqnarray}
&& \sigma\, \equiv\, \frac{n}{16}(7-p)^2(p-2)+ \frac{n}{16}(p-3)^2(8-p) ,\quad s\,\equiv \, n(7-p). \label{Eq:def_parameters1}
\end{eqnarray}
In order to take care of the divergent contribution to (\ref{Eq:area_Einst}), we set a cutoff at $r=\Lambda$. Then, the divergent part of the entropy is
\begin{equation}
S_{A,{\rm div}} \,=\,2\frac{\mathcal{V}}{4G_N\ell^{d-1}} \ell^{8-p-\sigma} \frac{\Lambda^{\sigma+1}}{\sigma+1},
\end{equation} which should be subtracted from (\ref{Eq:area_Einst}).

To perform the integrals above, it is convenient to redefine the radial coordinate
\be
u\,\equiv \, \frac{r_{\rm min}}{r},
\ee which leads to
\begin{equation}
S_{A,{\rm finite}}\,=\,2\frac{\mathcal{V}}{4G_N\ell^{d-1}} \ell^{8-p-\sigma} r_{\rm min}^{1+\sigma}  \int_{r_{\rm min}/\Lambda}^{1} du \frac{u^{-2-\sigma}}{\sqrt{1-u^\omega}} \frac{1}{\sqrt{1-\left( \frac{r_H}{r_{\rm min}}u\right)^{s}}} - S_{A,{\rm div}}, \label{Eq:area_u}
\end{equation} and
\begin{equation}
L\,=\,2 \ell^{2\xi}r_{\rm min}^{1-2\xi}\int_{0}^{1} du \, u^{-2+4\xi+\sigma}  \frac{1}{\sqrt{ 1 - u^{\omega} } \sqrt{1- \frac{r_H^s}{r_{\rm min}^s}u^s} }.  \label{Eq:L_u}
\end{equation} In the appendix, we include some analytic expressions that approximate the result in equations (\ref{Eq:area_u}) and (\ref{Eq:MI_Dp}).

Finally, we compute the area of the surface that interpolates between the two boundaries. This area corresponds to four times the area of a surface that divides the boundary in half and extends from one boundary to the black hole horizon. Defining $\tilde{u} \equiv \,u/r_H$,
\begin{eqnarray}
S_{A\cup B,{\rm finite}}=&&4\,\frac{\mathcal{V}}{4G_N\ell^{d-1}} \ell^{8-p-\sigma} r_H^{\sigma+1}\int_{r_H/\Lambda}^{1} d\tilde{u} \frac{\tilde{u}^{-2-\sigma}}{\sqrt{1-\tilde{u}^s}}  - 2\,S_{A,{\rm div}}, \label{Eq:area_AuB_Dp} \\
=&& 4\,\frac{\mathcal{V}}{4G_N \ell^{d-1}} \ell^{8-p-\sigma} r_H^{\sigma+1} \frac{\sqrt{\pi } \Gamma \left[-\frac{\sigma +1}{s}\right]}{s \Gamma \left[\frac{s-2 \sigma -2}{2 s}\right]} . \nonumber
\end{eqnarray}

Thus, the mutual information between $A$ and $B$ is given by
\be
I(A,B)=  \frac{\mathcal{V}\,\ell^{8-p-\sigma}}{G_N\ell^{d-1}}  r_{\rm min}^{1+\sigma}\left(  \int_{r_{\rm min}/\Lambda}^{1} du \frac{u^{-2-\sigma}}{\sqrt{1-u^\omega}} \frac{1}{\sqrt{1-\left( \frac{r_H}{r_{\rm min}}u\right)^{s}}}  - \int_{r_H/\Lambda}^{1} d\tilde{u} \frac{\tilde{u}^{-2-\sigma}}{\sqrt{1-\tilde{u}^s}} \right) . \label{Eq:MI_Dp}
\ee

\begin{figure}[t]
\centering
\hskip -0.12in
\includegraphics[width=6.7cm]{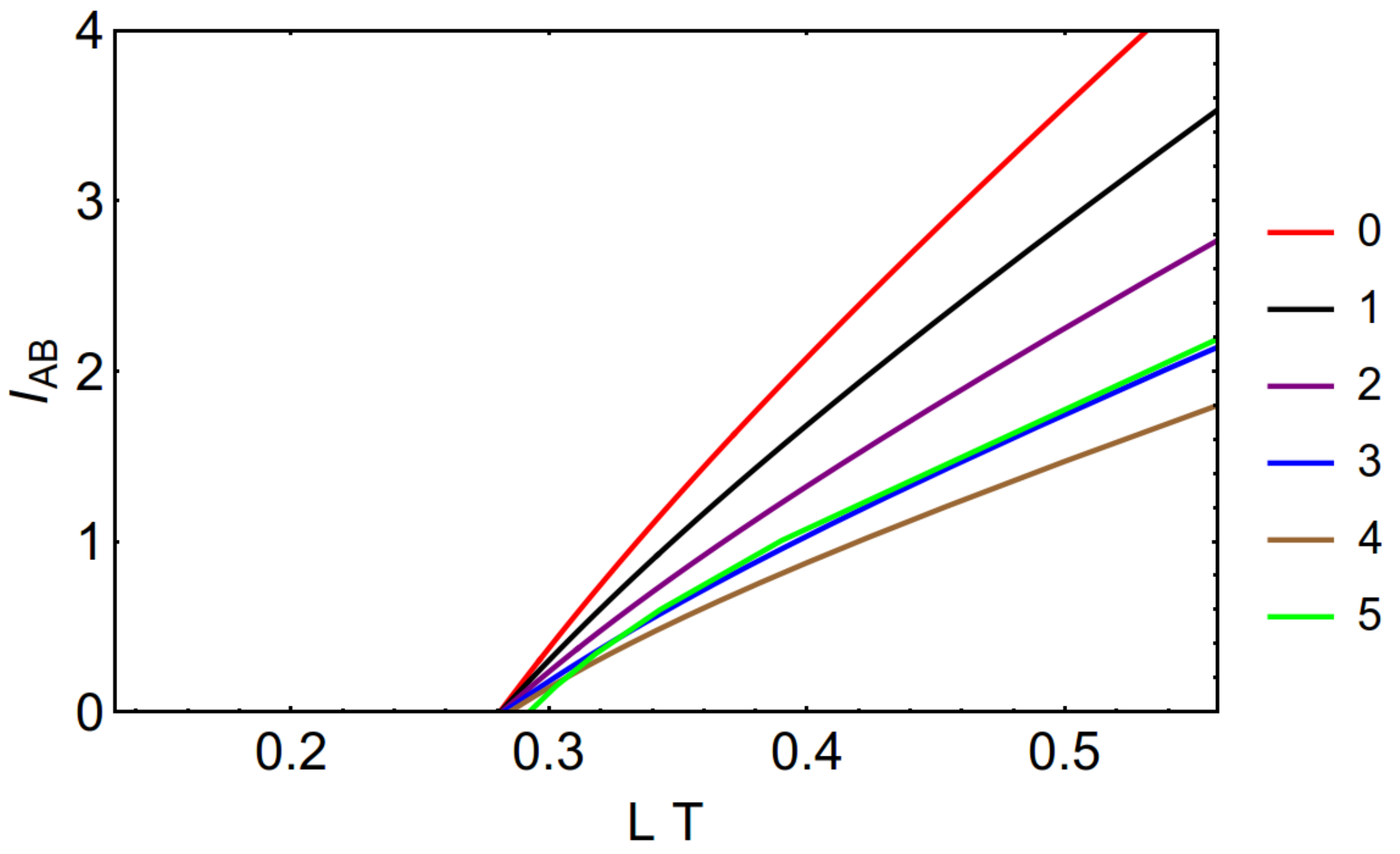}
\hskip 0.6in
\includegraphics[width=7.3cm]{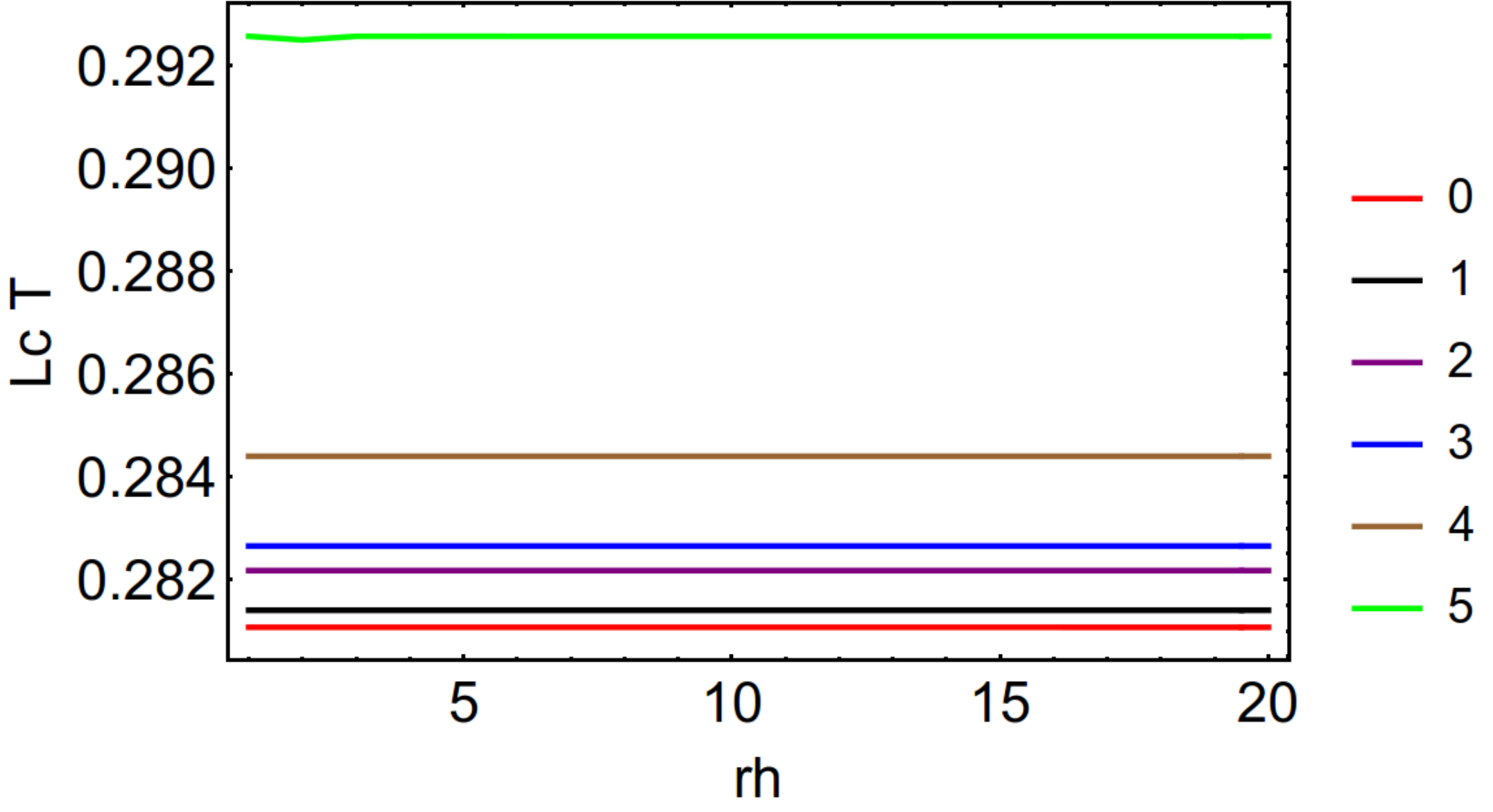}
\hskip 0.1in
\caption{Left: Examples of  $I(A,B)$ for $r_H/\ell\,=\,1$, for different values of $p$.
 Right: Critical values of $L$ for which $I(A,B)$ vanishes, in units of temperature. \label{Fig:Lcrit}}
\end{figure}

The mutual information defined above is non-zero only when the expression in first bracket is positive.
Mutual information between the regions in the two boundaries is relevant for values of $L>L_c$. The critical $L_c$, beyond which for which mutual information is non zero,
as a function of the horizon $r_H$ is depicted in Figure \ref{Fig:Lcrit}.

\subsection{Shock wave in black $Dp$ branes and scrambling time}\label{Sec:ScramblingTimeDp}
Let us now consider the back-reacted geometry due to a shock wave in the background of black $Dp$ branes. The analysis is a specialization of that in Section \ref{Sec:GenSWMetric}.
The back-reacted geometry is given by a shift $\alpha$ in $v$ coordinate and is  given by \eqref{alpha_genmetric},
\be
\alpha = \frac{c_2}{c_1} \frac{E \beta}{S_{BH}} e^{\frac{2 \pi}{\beta}\left(t_{w}-r_*(\Lambda )\right)},  \label{alpha_Dp}
\ee
where,
\bea
r_*(\Lambda ) &=& \frac{2 \ell }{n (p-5)} \left(\frac{\Lambda}{\ell} \right)^{\frac{n}{2} (p-5)},\,~~p \ne 5  \\
&=& \ell \log \left( \frac{\Lambda}{r_h}\right),\,~~ p=5,\nonumber \\
c_1 &=& r_h \frac{V_{d-1}'(r_h)}{V_{d-1}(r_h)} =\frac{\omega}{2} ,\nonumber\\
c_2 &=& r_h C(r_h,r_h)=e^{\psi(\frac{5-p}{2(7-p)})+\gamma + \log (n (7-p))},\,~~p \ne 5\,\nonumber\\
&=& 4 =c_1,\,~~ p=5; \nonumber
\eea
$\omega$ was defined in (\ref{Eq:Const-defs}), $\psi(x)$ is the Digamma function and $\gamma$ is Euler Constant. $E$ is the energy density of the perturbation and $S_{BH}$ is the black brane entropy density.

Following \eqref{ts_general_heuristic}, the scrambling time is given by setting $\alpha =1$ and $E\sim T$, which yields
\be
t_* = r_*(\Lambda) + \frac{\beta}{2 \pi} \log S_{BH} +  \frac{\beta}{2 \pi} \log \frac{c_1}{c_2},
\ee
where $c_1,\,c_2,\, r_*(\Lambda )$ are as given in previous equations. Notice that for $p<5$, $r_*(\Lambda )\to 0$ as $\Lambda \to \infty$ but for $p\ge 5$, $r_*(\Lambda )$ is divergent.
We will not try to draw conclusions in the case $p=5$, in the rest of the article, as the thermodynamics or holography in general is not very well defined for $D5$ brane~\cite{Barbon:2008sr}.

\subsection{Extremal surfaces for half plane in TFD}\label{Sec:Ext_Surf_HP_Dp}
We now compute EE using the bulk geometry given by Einstein frame metric \eqref{metricEF1} in the presence of a shock wave.  We  consider the case where the entangling region ($A_h/B_h$) at the left/right
boundary, which is half of the space parallel to the brane and filling up the transverse direction. Then due to symmetry the minimal surface corresponding
to the entangling region $A_h \cup B_h$, in the bulk, divides the  space parallel to $Dp$ brane in half,
and area minimization is reduced to a two dimensional problem. The surface is given as $r(t)$. Then the induced metric on the
bulk surface is
\be
ds_h^2 =( -f(r) +\frac{\dot r^2}{f(r)})dt^2+\left(\frac{r}{\ell}\right)^{\frac{n (7-p)^2}{8}} d\vec{y}_{\parallel,(p-1)}^2
+R_p^2 \left(\frac{r}{\ell}\right)^{\frac{n(p-3)^2}{8}} d\Omega_{8-p}^2 .
\ee
The area functional for the minimal surface is then given by
\be
\text{Area}= \mathcal{V}_{p-1} \Omega_{8-p} \frac{\ell^{8-p-m}}{n^{8-p}} \int dt ~ r^{m} \sqrt{-f+f^{-1} \dot r^2},
\ee
where, $m=\sigma+\frac{n}{16} (7-p)^2$ ($\sigma$ is defined in \eqref{Eq:def_parameters1}). The conserved quantity associated with $t$-translation symmetry,
\be
\gamma = \frac{-f r^{m}}{\sqrt{-f+f^{-1} \dot r^2}}=\sqrt{-f_0} r_0^{m},
\ee
where $r_0$ is defined as the coordinate for which $\dot r=0$. $r_0$ is assumed to be lying inside the horizon, thus,
$f_0=f(r_0)$ is negative. In the limit $r_0 \to r_h$, we have $\gamma \to 0$, and this should correspond to the limit $\alpha \to 0$
where the shock wave is absent.
\begin{figure}
\begin{center}
 \includegraphics[scale=.32]{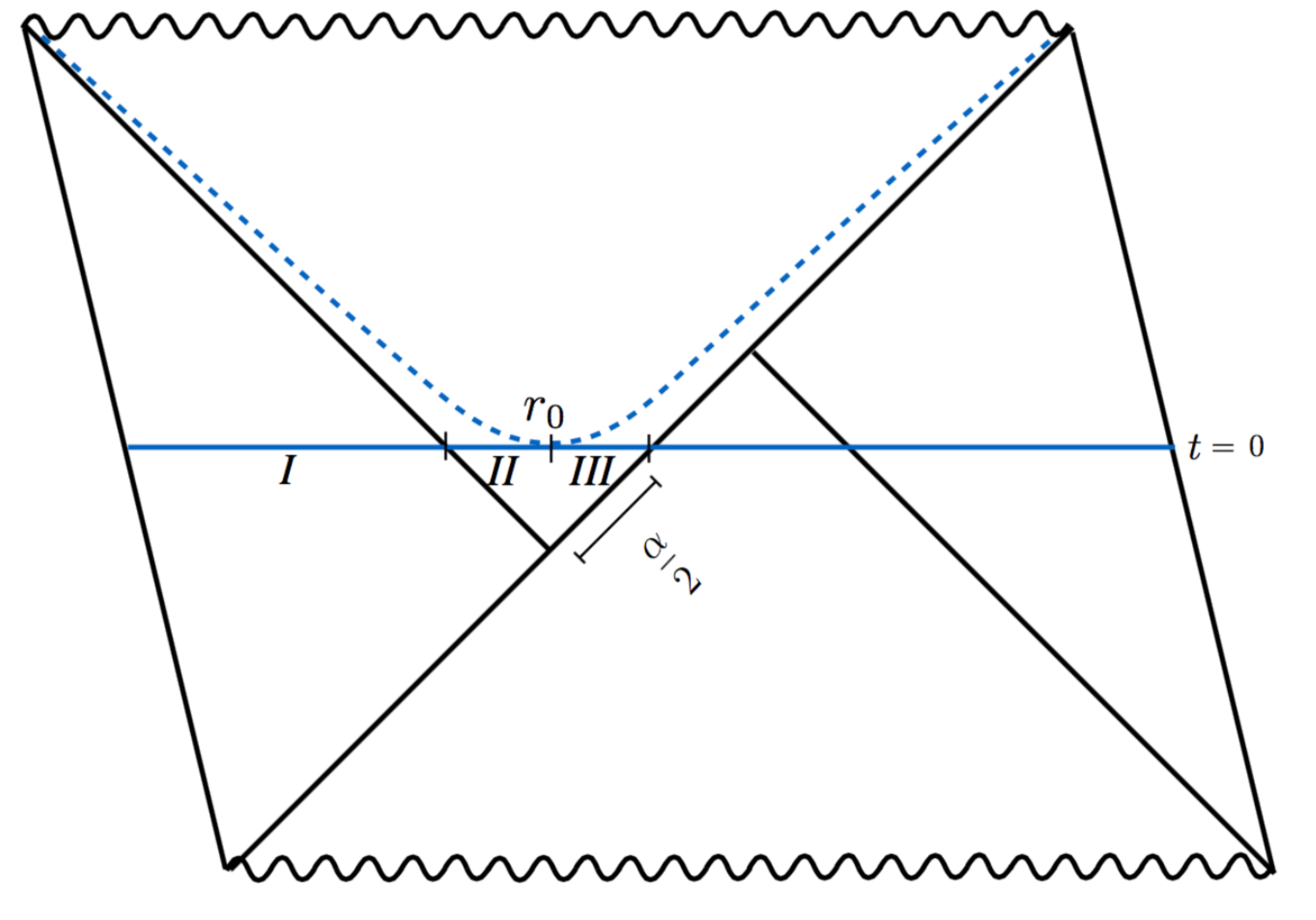}
 \caption{The minimal surface}\label{minimalsurf1}
 \end{center}
\end{figure}
The time coordinate $t$ as a function of radius $r$ along the extremal curve is given by,
\be
t(r)= \int \frac{dr}{f \sqrt{1+\gamma^{-2} f r^{2 m}}} .
\ee
The entanglement entropy is given by~\cite{Ryu:2006ef,Klebanov:2007ws},
\be
S_{A_h \cup B_h}= \frac{1}{4 G_{N}} \mathcal{V}_{p-1} \Omega_{8-p} \frac{\ell^{8-p-m}}{n^{8-p}} \int dr ~ r^{m} \frac{1}{\sqrt{f+\gamma^2 r^{-2m}}} .
\ee

Let us divide the minimal surface of the left half in three parts, as in Figure \ref{minimalsurf1}. The first segment goes from boundary at $(u_{\Lambda}, -u_{\Lambda})$
to $v=0$ ( at some value of $u$), the second from $v=0$ to $r=r_0$ (at some value of $t$), and third from $r=r_0$ to $u=0$. We can now compute
the area of each segment, and multiply the answer by two to get the total area. The second and third segments ($r=r_h$ to $r=r_0$ and back) manifestly
have same area. So the total area is given by twice the area of the first segment and $4$ times the area of the second segment. The entanglement entropy/ area of the minimal surface
is a function of $r_0$, which can be related to $\alpha$ as explained in the next paragraph.

\begin{figure}[t]
\begin{center}
\includegraphics[scale=.4]{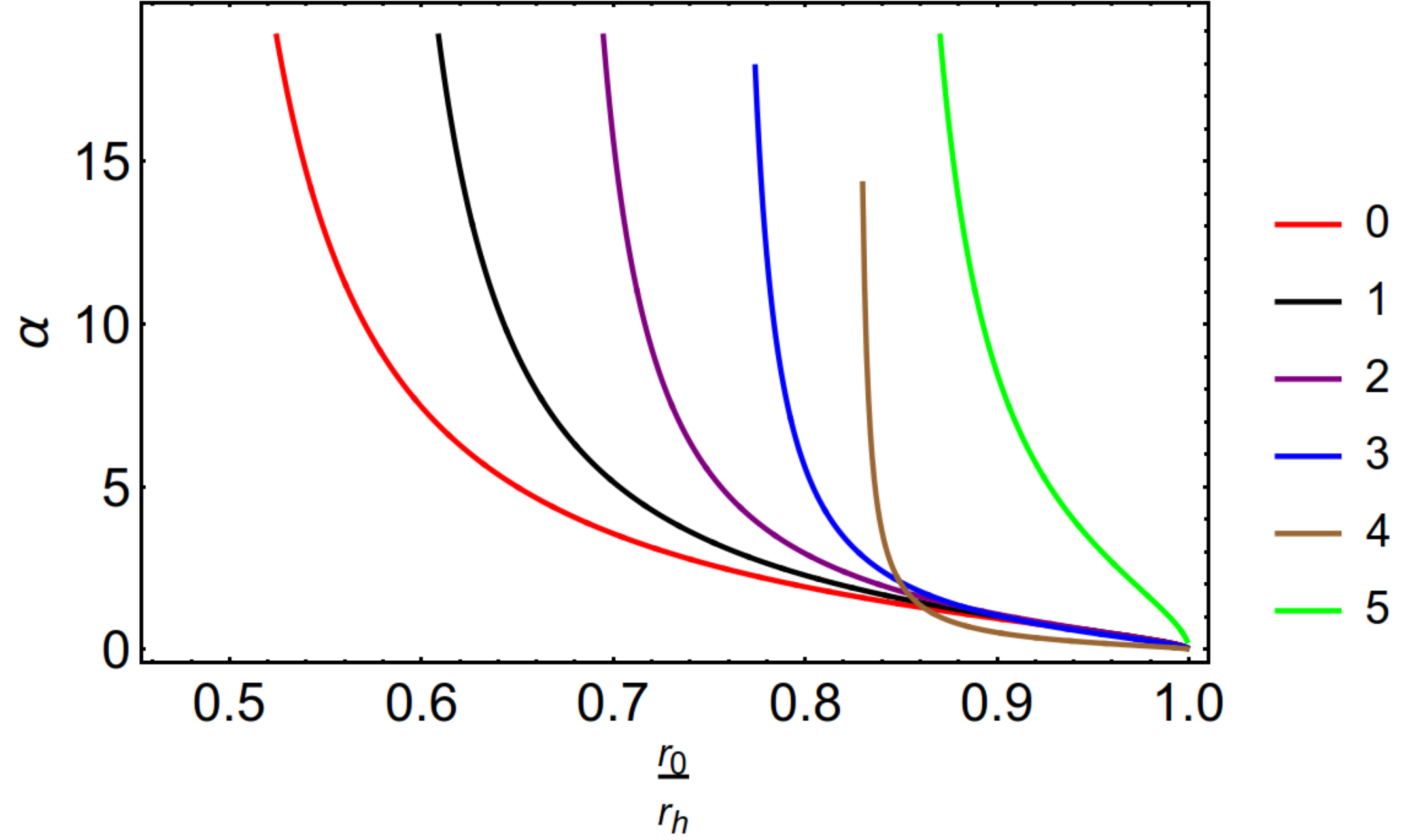}
\caption{$\alpha$ as a function of $\frac{r_0}{r_h}$ at any temperature.}\label{Fig:alpha_x0_Dp}
 \end{center}
\end{figure}

Following a similar analysis as that in \cite{Leichenauer:2014nxa}, we obtain a
relation between $\alpha$ and $r_0$ (see eq.(36) in \cite{Leichenauer:2014nxa}),
\be
\alpha = 2\frac{\bar u \bar v}{u_{\Lambda}} \exp(K_1+K_2+K_3)(r_0),  \label{Eq:alpha_r0_rel_Dp}
\ee
where,
\bea
K_1 &=& \frac{4 \pi}{\beta} \int_{\bar{r}}^{r_0} \frac{dr}{f}, \\
K_2 &=& \frac{2 \pi}{\beta} \int_{r_h}^{r_{\Lambda}} \frac{dr}{f} \left( 1-\frac{1}{\sqrt{1+\gamma^{-2} f r^{2m}}}\right), \nonumber \\
K_3 &=&\frac{4 \pi}{\beta} \int_{r_0}^{r_h} \frac{dr}{f} \left( 1-\frac{1}{\sqrt{1+\gamma^{-2} f r^{2m}}}\right), \nonumber
\eea
where $(\bar u,\bar v)$ is a reference point in the interior of the black brane \emph{i.e.} at some radial coordinate $r=\bar{r}<r_h$. If we choose $r_*(\bar r)=0$, then
$\bar u \bar v =1$ ~\cite{Leichenauer:2014nxa}. Figure \ref{Fig:alpha_x0_Dp}
shows numerical plots of dependence of $\alpha$ on $\frac{r_0}{r_h}$ for various $p$. We can show that this behavior is independent
of temperature for $\frac{\Lambda}{r_h}\gg 1$.

We can now use the relation between $\alpha$ and $r_0$, to get EE as a function of $\alpha$. The resultant EE has divergent contributions that are $\alpha$-independent. We can define a renormalized entanglement entropy as,
\be
S_{A_h \cup B_h}^{ren}(\alpha)= S_{A_h \cup B_h}(\alpha)- S_{A_h \cup B_h}(0).\label{Eq:RegEEHP-Dp}
\ee
These calculations can be performed analytically for $p=5$ as given in Appendix \ref{Appendix:AnalyticExtSurf-D5}. We use the results in Appendix\ref{Appendix:AnalyticExtSurf-D5}
to test our numerics.
\subsection{Mutual Information in shock wave black $Dp$ brane}\label{Sec:MI_SW_Dp}
Let us now consider the two strip-like regions $A$ and $B$ in the left/right boundary same as  that considered in Section \ref{Sec:MI_NonSW_Dp}.
We are again interested in the mutual information for this configuration, but now with a backreacted geometry due the presence of a shock wave.
$S_A+S_B$ is not affected due to the shock wave as the corresponding surface does not cross the horizon, so we can use the result in Section \ref{Sec:MI_NonSW_Dp}.
We can easily verify,
\be
I(A;B)(\alpha,L)=S_A+S_B-S_{A\cup B}= I^{NSW}(A;B)(L)-2 S_{A\cup B}^{reg}(\alpha),
\ee
where $I^{NSW}(A;B)$ is the mutual information computed in absence of the shock wave as computed in Section \ref{Sec:MI_NonSW_Dp}
and $S_{A\cup B}^{reg}(\alpha)$ is the regularized entanglement entropy for half-regions as computed in Section \ref{Sec:Ext_Surf_HP_Dp}.
Following the definition of mutual information, whenever the result is negative, it should be considered as zero. For a given length of
the strip $L$ in units of temperature, we can define $\alpha=\alpha_*$ where mutual information vanishes.
\begin{figure}
\centering
\hskip -0.12in
\includegraphics[width=7cm]{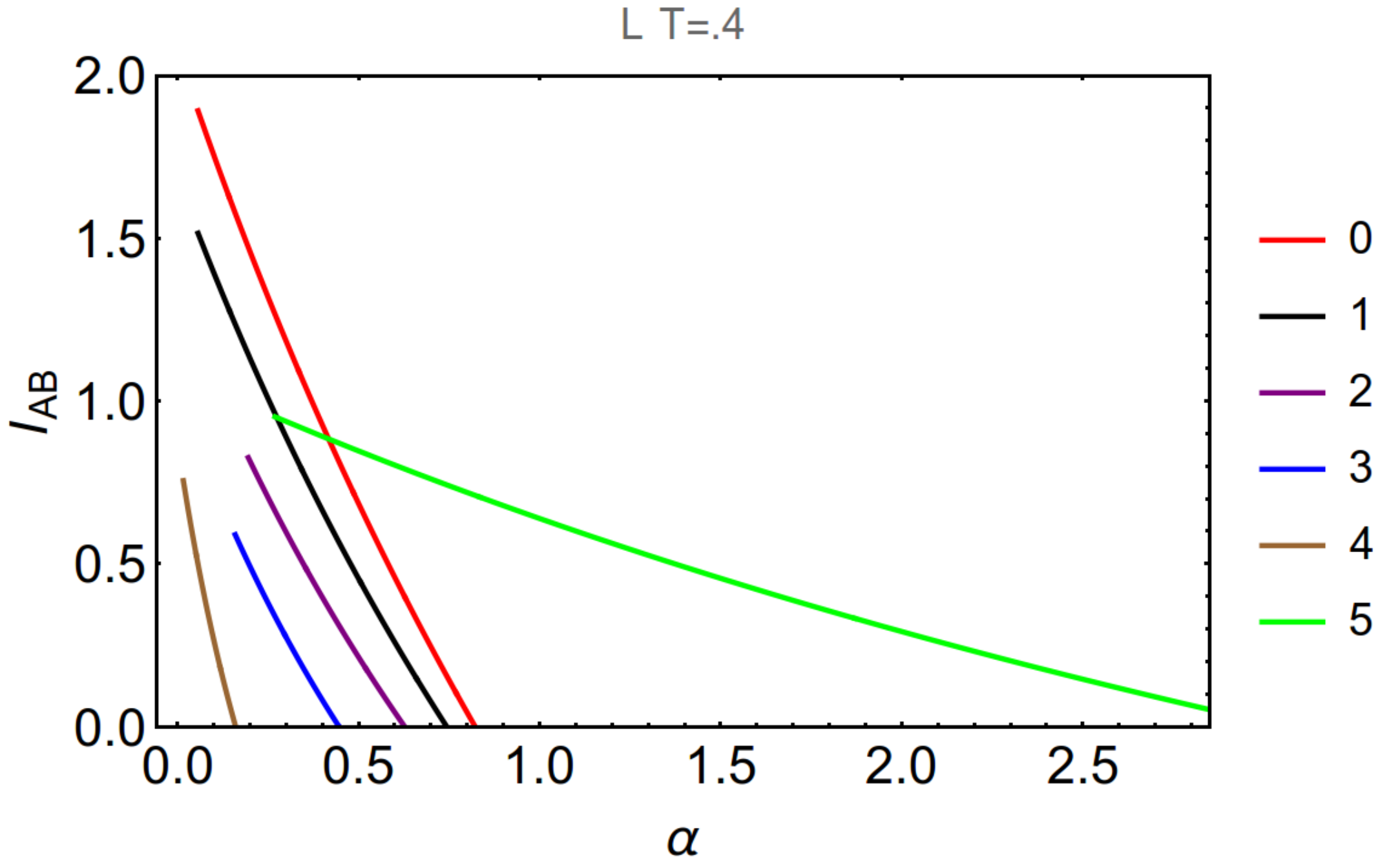}
\hskip 0.6in
\includegraphics[width=7cm]{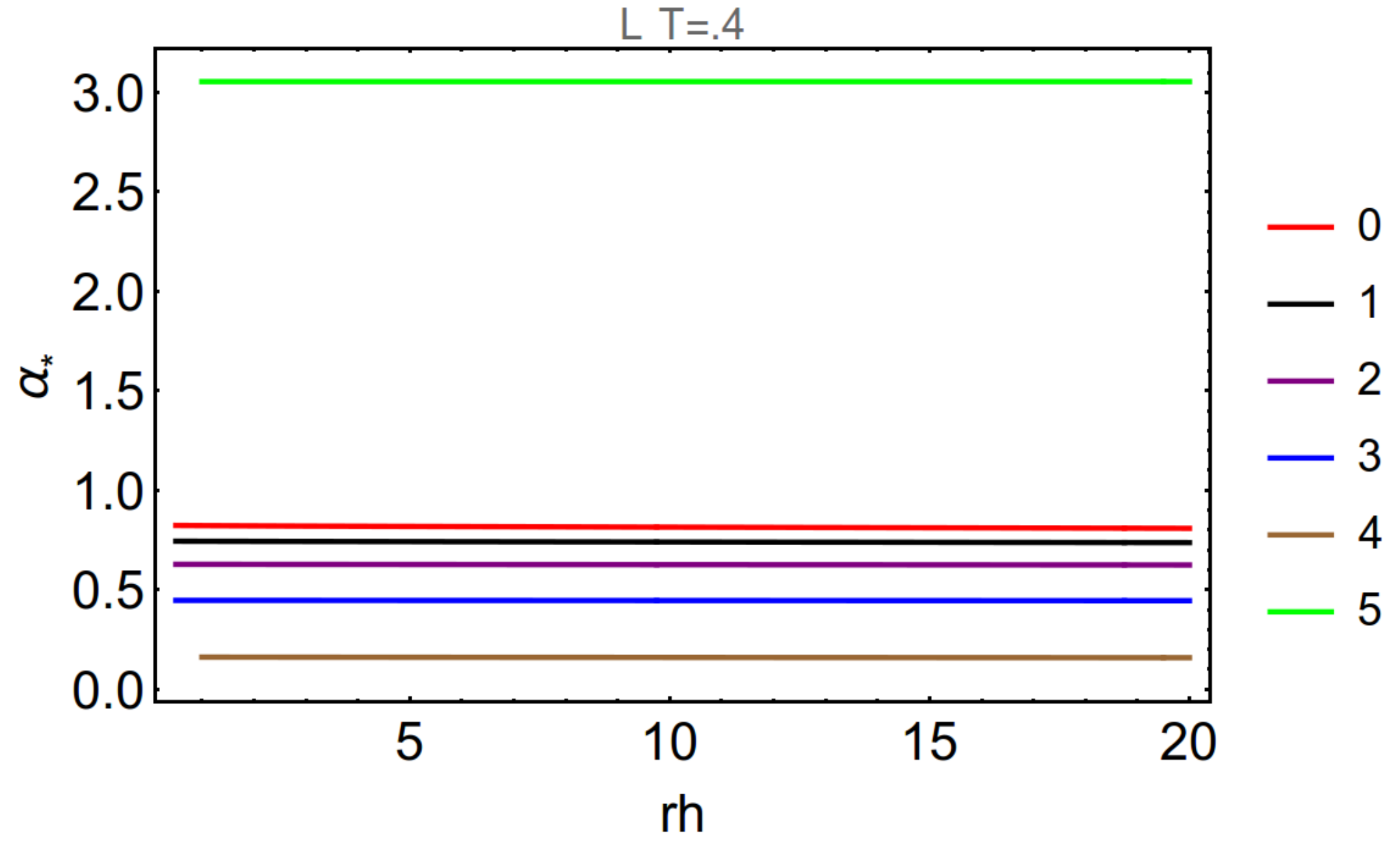}
\hskip 0.1in
\caption{Left: $I(A;B)$ as a function of $\alpha$ for various $p$ branes for $L= .4 \beta$.
Right: $\alpha_*$ as a function of horizon radius, in units of $\ell$.} \label{Fig:MI_Dp_Shock}
\end{figure}

\begin{figure}{t}
\begin{center}
 \includegraphics[scale=.35]{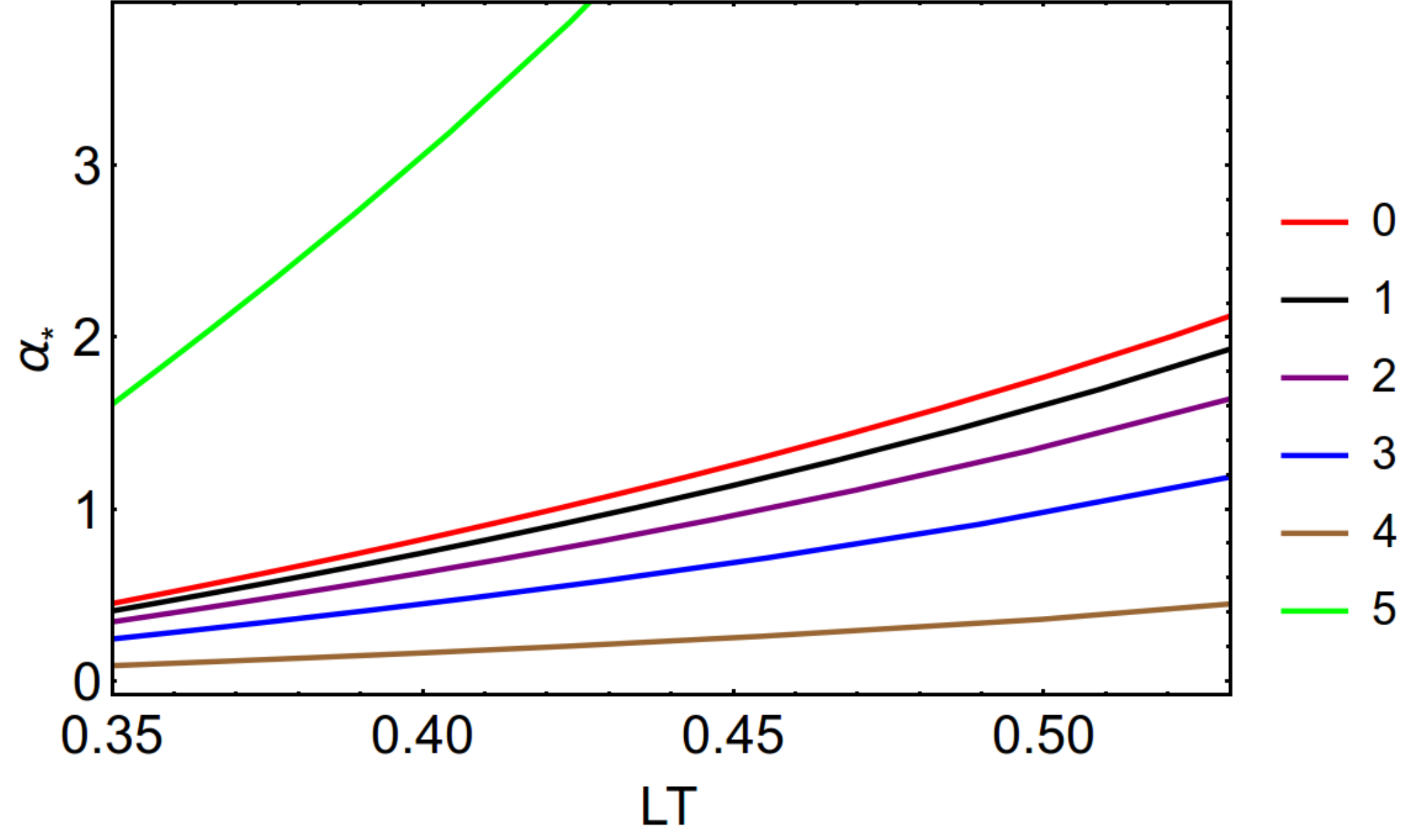}
 \caption{$\alpha_*$ as a function of boundary strip length  in units of temperature ($LT$).}\label{Fig:alpha_LT_Dp_Shock}
\end{center}
\end{figure}
The behavior of mutual information as function of $\alpha$
and $\alpha_*$ as a function of horizon radius is represented in Figure \ref{Fig:MI_Dp_Shock}. Note that we are only interested in the behavior of mutual
information as a function of $\alpha$ and not the value itself, so we have neglected overall constants. From Figure \ref{Fig:MI_Dp_Shock}, it is  clear that
$\alpha_*$ is independent of temperature or $r_h$, as long as we keep the length of the boundary strip fixed in units of temperature. The value of $\alpha_*$
decreases as we increase $p$ from $0$ to $4$ for a given length of the boundary strip  in units of temperature. The behavior for $p=5$ does not match this pattern.
Figure \ref{Fig:alpha_LT_Dp_Shock} shows the behavior of $\alpha_*$ as a function of the length of the boundary strip  in units of temperature. We now translate
the behavior of $\alpha_*$ to that of $t_*$ using \eqref{alpha_Dp} (with $E \sim T$),
\be
t_*=r_*(\Lambda)+\frac{\beta}{2\pi} \left( \log S_{BH}(\beta,p,\lambda, N) + \log \frac{c_1}{c_2}(p) + \log \alpha_*(\frac{L}{\beta},p)\right), ~~p<5.
\ee
We know from \eqref{Eq:Dp_Entropy},
\be
S_{BH}=N^2 s(p,\lambda,\beta),
\ee
where the function $s(p,\lambda,\beta)$ can be read off from \eqref{Eq:Dp_Entropy} and is some $\mathcal{O}(N^0)$ number, similar to the terms $\log \frac{c_1}{c_2}(p),
\log \alpha_*(\frac{L}{\beta},p)$. 
The term $r_*(\Lambda)$ is small for $\Lambda \gg1$ and $p<5$. 



\section{A black hole in asymptotically Lifshitz solutions}\label{Sec:Nonrelativistic}


\subsection{Bulk metric}\label{Sec:Lif_metric}
In this section we shift our interest to chaos in non-relativistic field theories. We are particularly focusing on non-relativistic
scale invariant field theories which arise as a fixed point description of various condensed matter systems. The scale invariance (at Lifshitz Fixed Point)
is given by,
\be
t \to \lambda^z t ~~~;~~~~ x \to  \lambda \,x ~~\text{for}~~z \ne 1.
\ee
In contrast to conformal fixed points, which correspond to $z=1$. $z$ is called dynamical critical exponent.
This anisotropic scaling of space and time makes these theories explicitly
non-relativistic. Holographic duals of such theories were proposed in \cite{Kachru:2008yh}, which were later generalized in \cite{Taylor:2008tg, Balasubramanian:2009rx}.
These dual gravity solutions are generically known as Lifshitz Solutions.
For a more recent review on this topic see ~\cite{Taylor:2015glc}. In particular, thermalization in these non-relativistic systems was studied holographically in
~\cite{Alishahiha:2014cwa, Fonda:2014ula}.

Now we look at an example of a black hole in an asymptotically Lifshitz background found in \cite{Balasubramanian:2009rx}.
In that work, the authors consider a $2+1$ boundary.
The bulk geometry is a solution to the equations of motion derived from the action
\be
S\,=\,\frac{1}{2} \int d^4x \,(R-2\Lambda)\,-\,\int d^4x\,\left(\frac{e^{2\Phi}}{4}F^2+\frac{m^2}{2}A^2+\left(e^{2\Phi}-1 \right) \right),
\ee where $A$ is a vector field and $F$ is the respective strength form. The solution to the equations of motion is given by

\begin{eqnarray}
&& ds^2\,=\,-f_0(\rho) \left(\frac{\rho}{\ell}\right)^{2z}dt^2\,+\,\left(\frac{\rho}{\ell}\right)^2d\vec{x}^2\,+\, \frac{d\rho^2}{\left(\frac{\rho}{\ell}\right)^2f_0(\rho)},\quad  f_0(\rho)\,=\,1-\frac{\rho_h^2}{\rho^2} \label{Eq:Lifshitz0} \\
&& A\,=\,\frac{1}{\ell} f_0(\rho) \left(\frac{\rho}{\ell}\right)^2 dt,\qquad \Phi\,=\,-\frac{1}{2} \log \left( 1-\frac{\rho_h^2}{\rho^2} \right).
\end{eqnarray}

The surface gravity of the black hole in (\ref{Eq:Lifshitz0}) leads to the temperature
\be
T\,=\,\frac{\rho_h^{z}}{2\pi \ell^{z+1}}.
\ee
The metric (\ref{Eq:Lifshitz0}) does not have the property $g_{tt}g_{\rho\rho} = -1$, which would facilitate the definition of the Kruskal coordinates as in previous examples. However, we can redefine the radial coordinate
\be
z^2 \left(\frac{r}{\ell}\right)^2 \, \equiv \,\left( \frac{\rho}{\ell}\right)^{2z},
\ee which leads to the metric
\be
ds^2\,=\,-f(r)dt^2\,+\, \frac{dr^2}{f(r)}\,+\,\left(\frac{r}{\ell}\right)^{2/z}d\vec{x}^2,\quad  f(r)\,=\,z^2\left(\frac{r}{\ell}\right)^2\left(1-\left(\frac{r_h}{r}\right)^{\frac{2}{z}} \right), \label{Eq:Lifshitz1}
\ee
which is equipped with a temperature
\be
T\,=\,\frac{r_h \,z}{2\pi \ell^2}.
\ee
and entropy,
\be
S_{BH}=\frac{V_2}{4 G^{(4)}_{N}} \left( \frac{2 \pi \ell}{z}\right)^{\frac{2}{z}} T^{\frac{2}{z}}.
\ee
Here we have rescaled the coordinates $\vec{x}=\{x_1,x_2\}$ by $z^{1/z}$.

\subsection{Mutual Information}\label{Sec:MI_NonSW_Lif}

Now we compute again the mutual information between two strips $A$ and $B$, contained in the left and right side of the geometry respectively.
The strips are defined as $0<x_1<L$ and $0<x_2<\mathcal{V}$ ($\mathcal{V} \to \infty$).
We use the metric given by (\ref{Eq:Lifshitz1}).

The entanglement entropy of one strip, $S_A$ or $S_B$, with width $L$ is given by the area

\begin{equation}
{\rm Area}\,=\,2\frac{ \mathcal{V}}{z}  \int_{r_{\rm min}}^{\infty}dr \frac{1}{\left( \frac{r}{\ell} \right)^{1-\frac{1}{z}} \sqrt{1-\left(\frac{r_H}{r}\right)^\frac{2}{z}} } \frac{1}{\sqrt{1-\left( \frac{r_{\rm min}}{r}\right)^{\frac{4}{z}}}}, \label{Eq:area_Lif1}
\end{equation}

In order to take care of the divergent contribution to (\ref{Eq:area_Lif1}), we set a cutoff at $r=\Lambda$. Then, the divergent part of the area is
\begin{equation}
{\rm Area}_{\rm div} \,=\,2 \mathcal{V} \ell \left( \frac{\Lambda}{\ell}\right)^{\frac{1}{z}}.
\end{equation}
$r_{\rm min}$ and the width $L$ are related by
\begin{equation}
L\,=\,\frac{2}{z}\int_{r_{\rm min}}^{\infty} dr \,   \frac{1}{\left(\frac{r}{\ell}\right)^{1+\frac{1}{z}}\sqrt{1-\left(\frac{r_H}{r}\right)^{\frac{2}{z}}} \sqrt{\left( \frac{r}{r_{\rm min}}\right)^{\frac{4}{z}} -1 }}.  \label{Eq:L_Lif1}
\end{equation}

The finite part of the area is defined as,
\begin{equation}
 {\rm Area}_{\rm finite}={\rm Area}-{\rm Area}_{\rm div}
\end{equation}

Now, we compute the area of the surface that interpolates between the two boundaries. This area corresponds to four times the area of a surface that divides the boundary in half and extends from one boundary to the black hole. Defining $\tilde{u} \equiv \,u/r_H$,

\be
{\rm Area}_{A \cup B,\rm finite}= 4\,\frac{ \mathcal{V}}{z} \int_{r_h}^{\infty} dr \, \frac{1}{\left(\frac{r}{\ell}\right)^{1+\frac{1}{z}}\sqrt{1-\left(\frac{r_H}{r}\right)^{\frac{2}{z}}}}  - 2\,{\rm Area}_{\rm div}, \label{Eq:area_AuB_Lif} \\
\ee

Thus, the mutual information between $A$ and $B$ is given by
\be
I(A,B)\,\equiv\, S_A+S_B-S_{A\cup B}=\frac{1}{4 G_N^{(4)}} \left(2 \, {\rm Area}_{\rm finite} -{\rm Area}_{A \cup B, \rm finite}\right) .
\ee

\begin{figure}
\centering
\hskip -0.12in
\includegraphics[width=6.3cm]{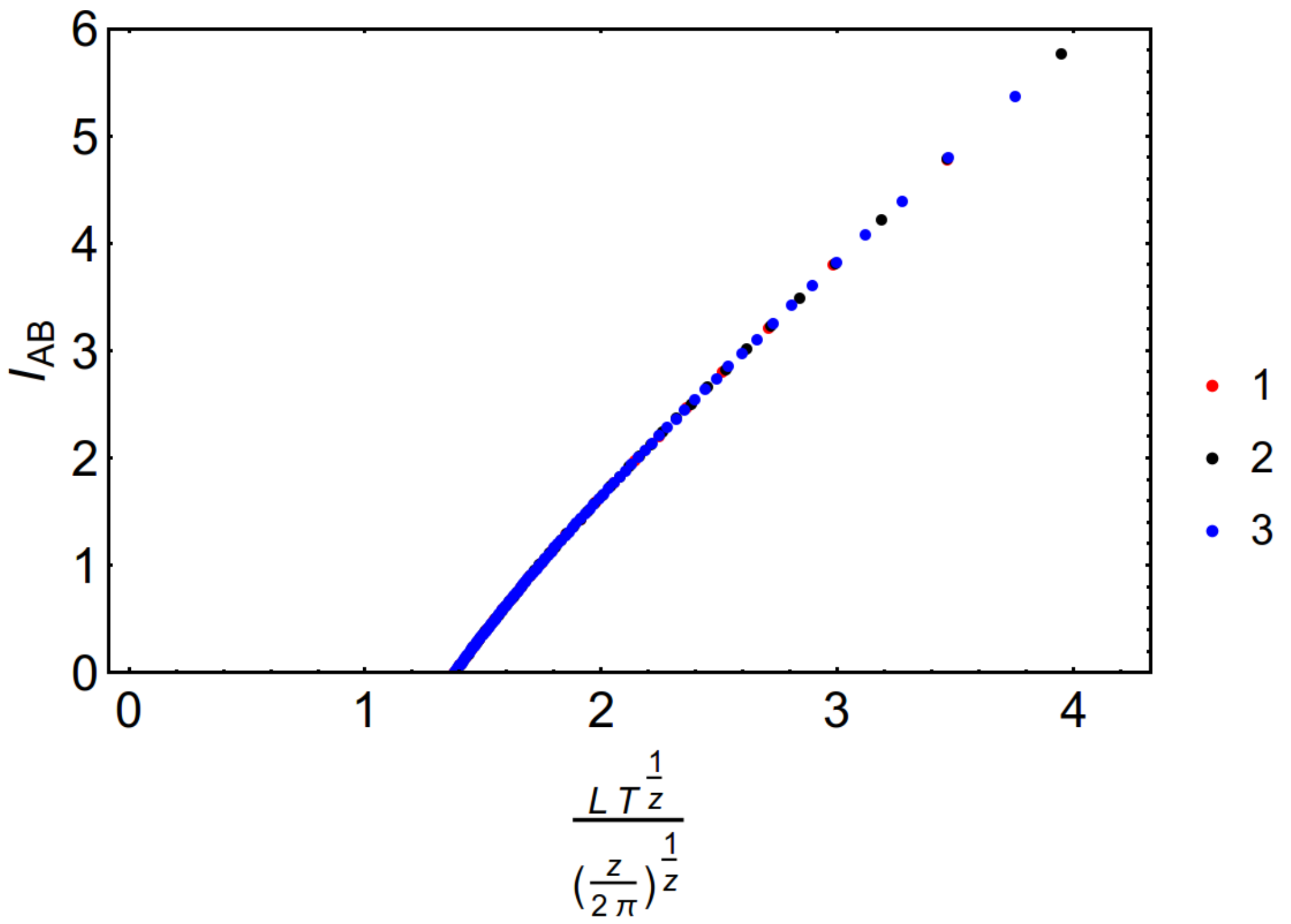}
\hskip 0.4in
\includegraphics[width=8cm]{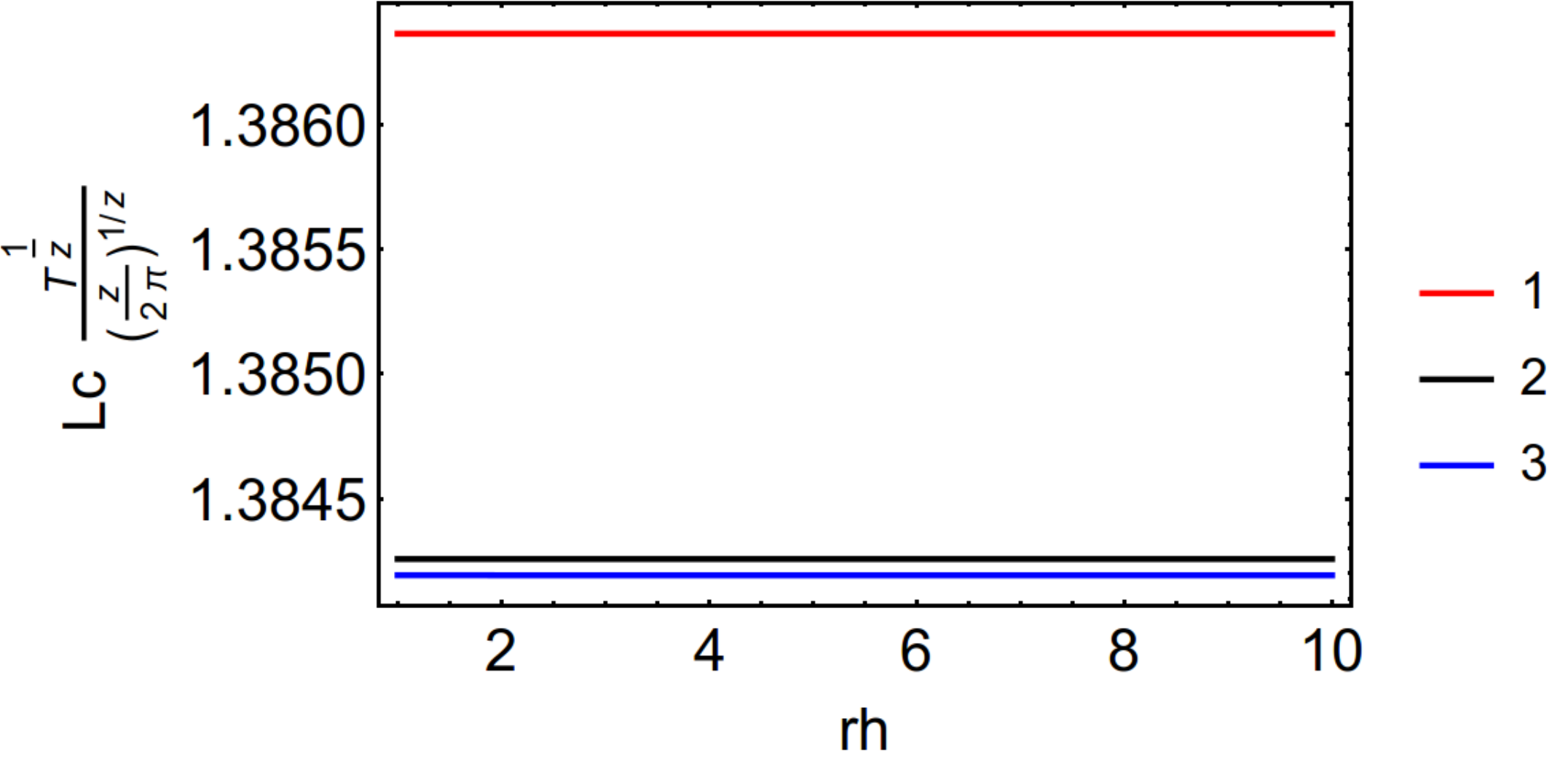}
\vskip 0.2in
\caption{Left: Examples of  $I(A,B)$ for $r_H/\ell\,=\,1$, for different values of $z$.
 Right: Critical values of $L$ for which $I(A,B)$ vanishes, in units of temperature. \label{Fig:Lcrit_Lif}}
\end{figure}

The critical $L_c$, for which mutual information vanishes, as a function of the horizon $r_H$ is depicted in Figure \ref{Fig:Lcrit_Lif}.

\subsection{Shock wave in Lifshitz black hole and scrambling time}\label{Sec:ScramblingTime_Lif}
Let us now consider the back-reacted geometry due to a shock wave in the background of Lifshitz black-hole.
The analysis is a specialization of Section \ref{Sec:GenSWMetric}.
The back-reacted geometry is given by a shift $\alpha$ in $v$ coordinate and is  given by \eqref{alpha_genmetric},
\be
\alpha = \frac{c_2}{c_1} \frac{E \beta}{S_{BH}} e^{\frac{2\pi}{\beta}\left(t_{w}-r_*(\Lambda )\right)},
\ee
where,
\bea
r_*(\Lambda ) &=& -\frac{\ell^2}{z^2 \Lambda},   \\
c_1 &=& r_h \frac{V_{d-1}'(r_h)}{V_{d-1}(r_h)} =\frac{1}{z}, \nonumber\\
c_2 &=& r_h C(r_h,r_h)=\frac{2}{z}e^{\psi(\frac{z}{2})+\gamma }, \nonumber
\eea
where $\psi(x)$ is the Digamma function and $\gamma$ is Euler Constant.

The heuristic scrambling time is given by setting $\alpha =1$ and $E=T$ is given by \eqref{ts_general_heuristic},
\be
t_* = r_*(\Lambda) + \frac{\beta}{2 \pi} \log S_{BH} +  \frac{\beta}{2 \pi} \log \frac{c_1}{c_2},
\ee
where $c_1,c_2, r_*(\Lambda )$ is as given in previous equations.

\subsection{Extremal surfaces for half plane in TFD}\label{Sec:Ext_Surf_HP_Lif}
In this section, we analyze the extremal surface that extends between the two boundaries of the geometry \eqref{Eq:Lifshitz1} in the presence of a shock wave. We consider the case where the entangling region ($A_h/B_h$), at the left/right
boundary respectively, is half of the space parallel to the brane and fills up the transverse direction.
This analysis is similar to that done in Section \ref{Sec:Ext_Surf_HP_Dp}.

\begin{figure}[h]
\begin{center}
\includegraphics[scale=.35]{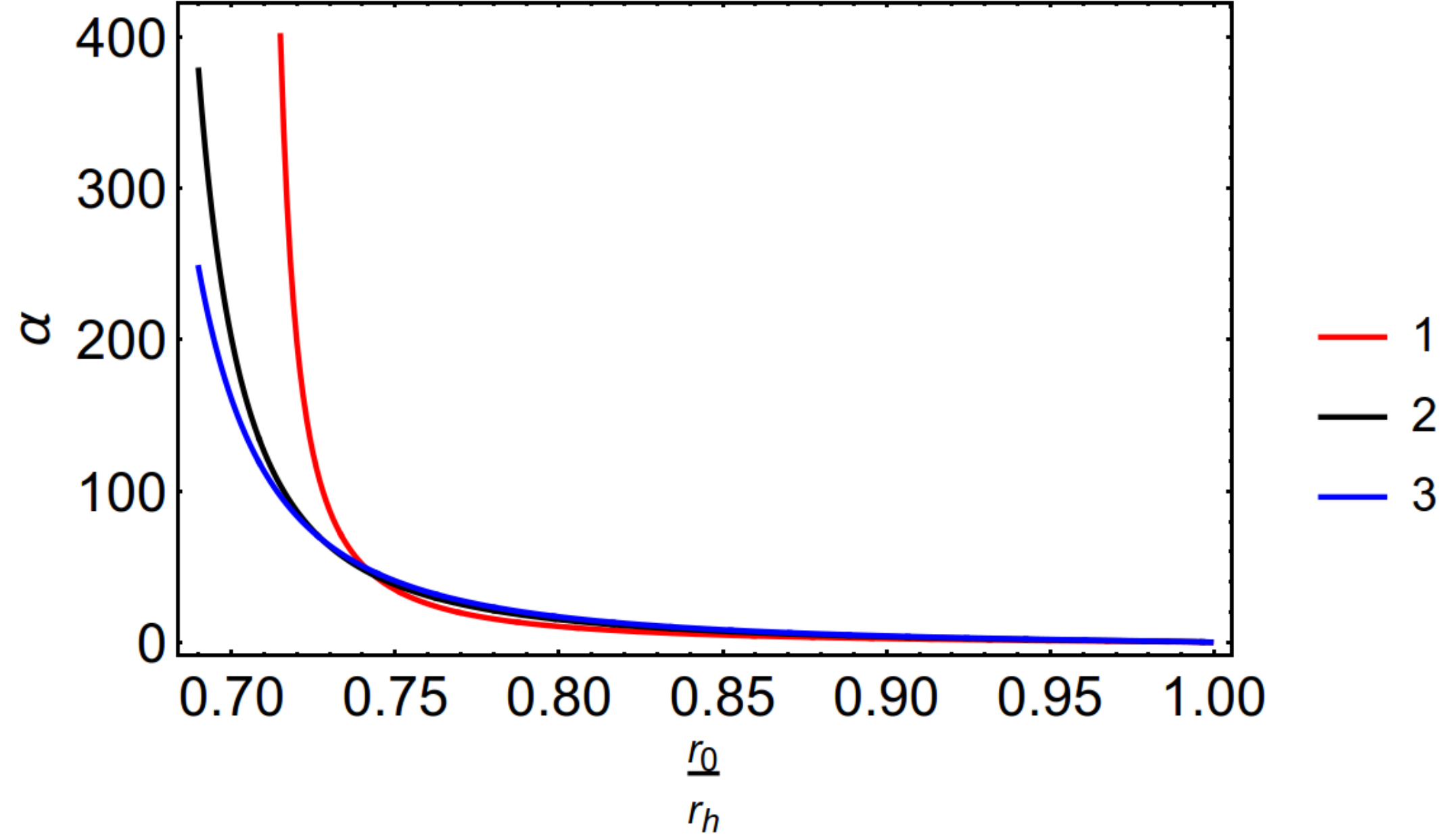}
\caption{$\alpha$ as a function of $\frac{r_0}{r_h}$ at any temperature.}\label{Fig:alpha_x0_Lif}
\end{center}
\end{figure}

The time coordinate $t$ as a function of radius $r$ along the extremal curve is now given by,
\be
t(r)= \int \frac{dr}{f \sqrt{1+\gamma^{-2} f r^{2 m}}} ~,~~~ m =\frac{1}{z},
\ee
where,
\be
\gamma =\sqrt{-f(r_0)} r_0^{m}
\ee
 and $r=r_0 (<r_h)$ is the deepest the surface penetrates the region beyond the horizon, as described in Section \ref{Sec:Ext_Surf_HP_Dp}.
The entanglement entropy is given by,
\be
S_{A_h \cup B_h}= \frac{1}{4 G_{N}} \mathcal{V} \int dr ~ r^{m} \frac{1}{\sqrt{f+\gamma^2 r^{-2m}}}
\ee
The entanglement entropy is a function of $r_0$, which can be related to $\alpha$, the shift in the shock wave geometry,
similarly to Equation (\ref{Eq:alpha_r0_rel_Dp}).
Figure \ref{Fig:alpha_x0_Lif} shows a numerical plots of dependence of $\alpha$ on $\frac{r_0}{r_h}$ for various $z$. We can show that this behavior is independent
of temperature for $\left(\frac{\Lambda}{r_h}\right)^{1/z}\gg 1$.

\subsection{Mutual Information in shock wave Lifshitz Black Hole}\label{Sec:MI_SW_Lif}
\begin{figure}
\centering
\hskip -0.12in
\includegraphics[width=7cm]{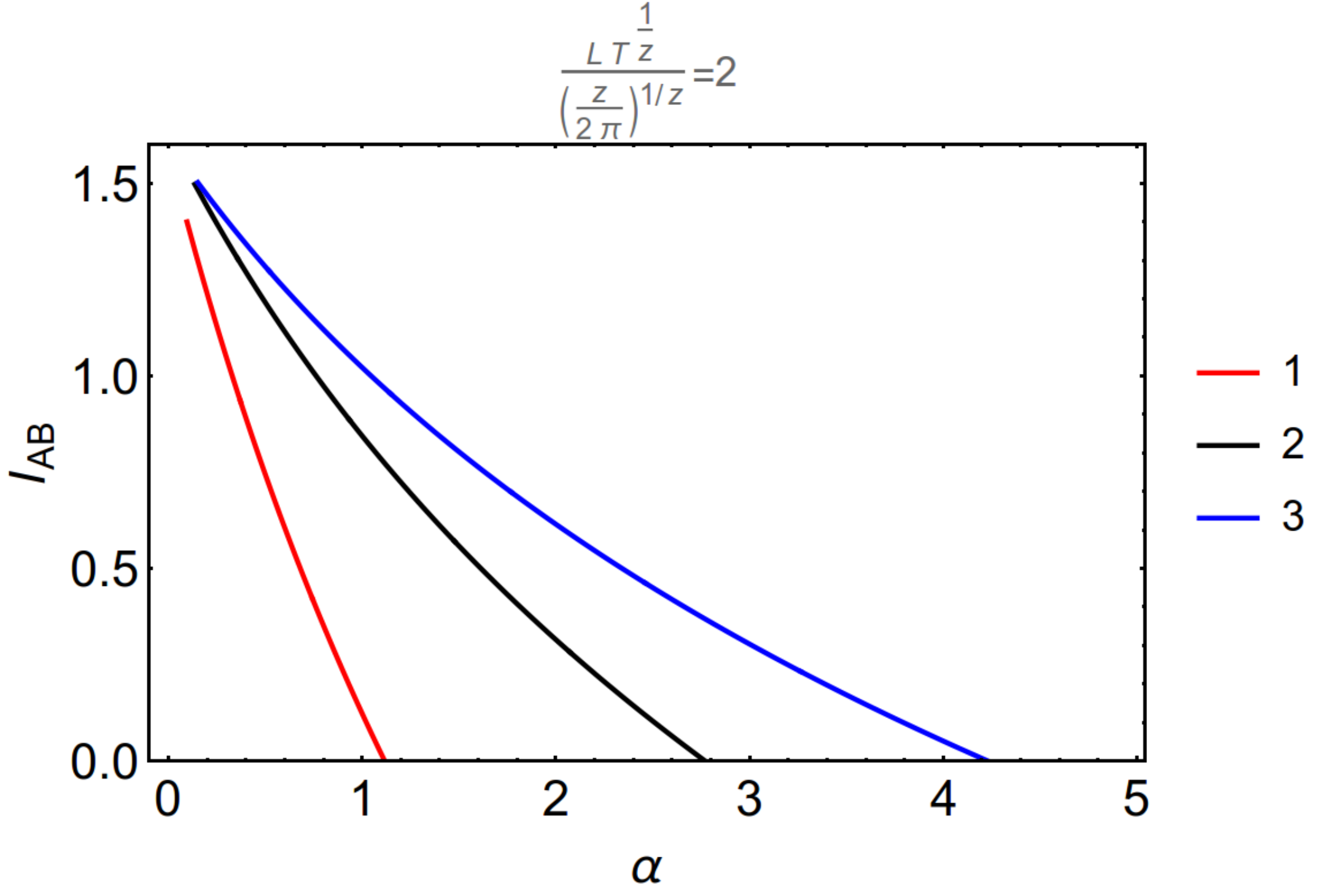}
\hskip 0.6in
\includegraphics[width=7cm]{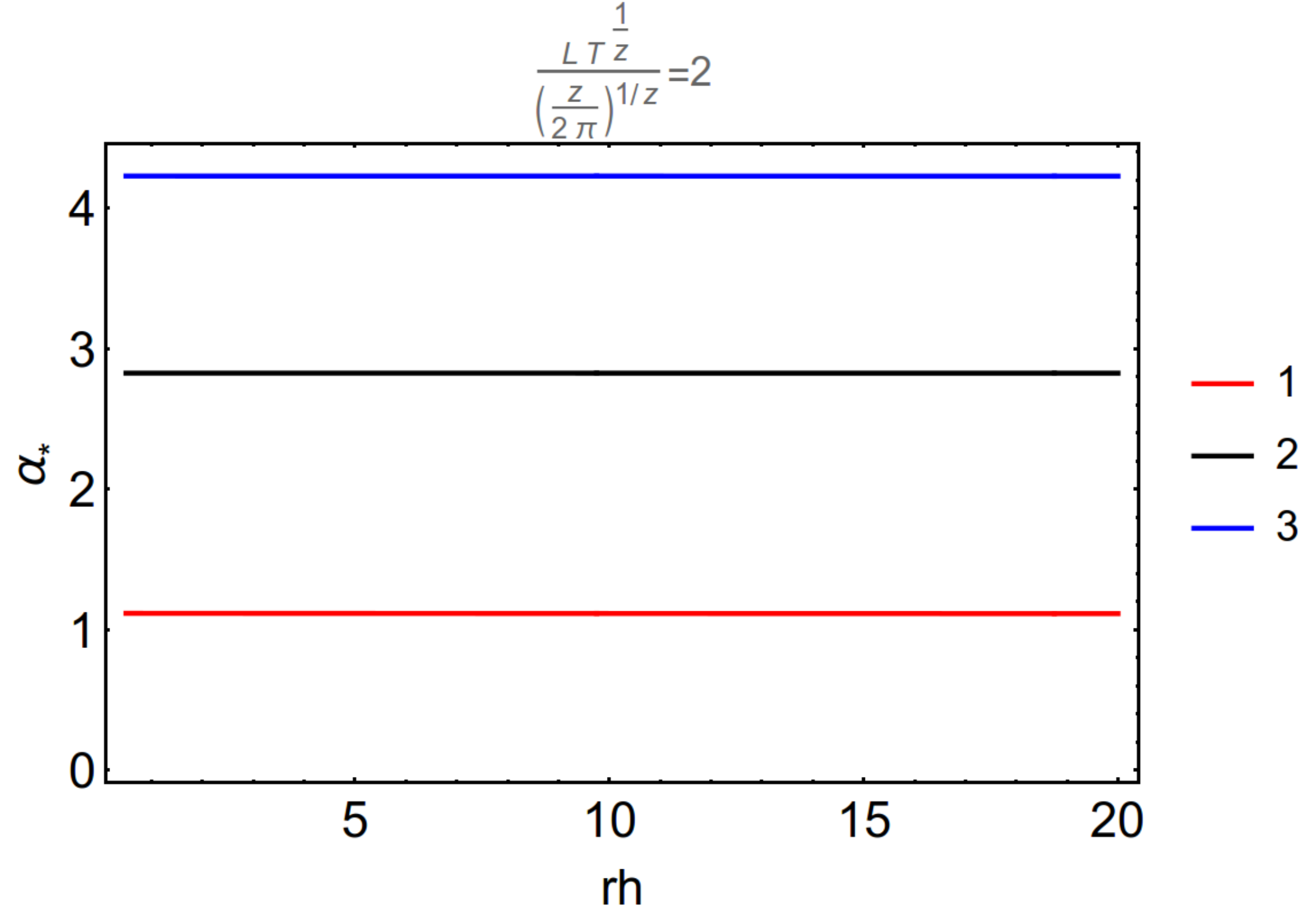}
\hskip 0.1in
\caption{Left: $I(A;B)$ as a function of $\alpha$ for various $z$ for $\frac{L T^\frac{1}{z}}{\left( \frac{z}{2\pi}\right)^\frac{1}{z}}=2$.
Right: $\alpha_*$  as a function of horizon radius, in units of $\ell$.}\label{Fig:IAB_alpha_Lif}
\end{figure}

\begin{figure}
\begin{center}
 \includegraphics[scale=.35]{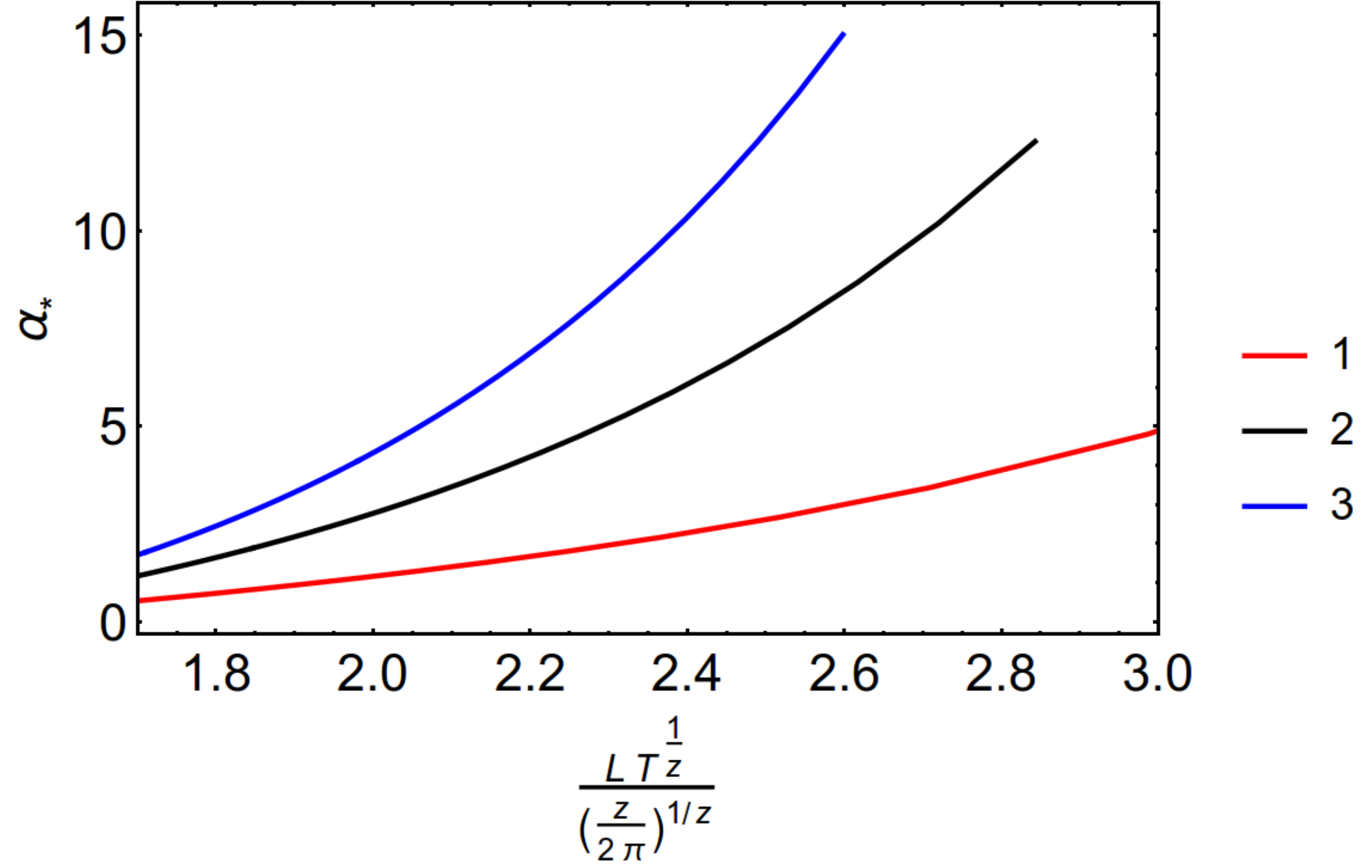}
 \caption{$\alpha_*$ as a function of length of the boundary strip fixed in units of temperature for various $z$.}\label{Fig:alphastar_LT_Lif}
 \end{center}
\end{figure}

We now calculate mutual information in the Lifshitz Black Hole in presence of a shock wave, for two-strip like regions defined in Section \ref{Sec:MI_NonSW_Lif}.
We follow the calculation in Section \ref{Sec:MI_SW_Dp}. Figure \ref{Fig:IAB_alpha_Lif} shows the behavior of mutual information for
various $z$ as function of $\alpha$ for a fixed $L$ in units of temperature (\emph{i.e.} fixed $L(2\pi T/z)^{1/z}$).
$\alpha_*$ gives the value of $\alpha$ for which the mutual information vanishes.
We can easily notice that $\alpha_*$ increases with $z$ for length of the boundary strip fixed in units of temperature. Also $\alpha_*$ is independent of $r_h$
or temperature for a given $z$ and length of the boundary strip fixed in units of temperature. Figure \ref{Fig:alphastar_LT_Lif} shows variation of $\alpha_*$
with the length of the boundary strip fixed in units of temperature.

\section{Gauss-Bonnet gravity}\label{Sec:HD_Theories}


\subsection{Black hole geometry in Gauss-Bonnet gravity}\label{Sec:HD_metric}

Finally, we would like to include in our analysis higher curvature corrections to the conformal case studied in \cite{Leichenauer:2014nxa}. Specifically, we focus on the planar black hole solution in Gauss-Bonnet gravity. The Gauss-Bonnet correction to the action for the  $5$-dimensional bulk in is given by \cite{Cai:2001dz,Myers:2010ru,deBoer:2011wk,Myers:2012ed}
\begin{eqnarray}
&& S_{\rm grav} = \frac{1}{16 \pi G_N} \int d^5x \sqrt{-g} \left( R + \frac{12}{L^2} + \frac{\lambda_{\rm GB} L^2}{2} \mathcal{X}_4\right)  \ ,  \label{GBaction}\\
&& \mathcal{X}_4 = R_{\mu\nu\rho\sigma} R^{\mu\nu\rho\sigma} - 4 R_{\mu\nu}R^{\mu\nu} + R^2,
\end{eqnarray}
where $R$ denotes the Ricci-scalar. The equations of motion that follow from this action are given by
\begin{eqnarray}
&& R_{\mu\nu} - \frac{1}{2} g_{\mu\nu} \left( R + \frac{12}{L^2} + \frac{\lambda_{\rm GB} L^2}{2} \mathcal{X}_4 \right) +  \lambda_{\rm GB} L^2 \mathcal{H}_{\mu\nu} = 0 \ , \label{Eq:eomGB} \\
&& \mathcal{H} = R_{\mu\rho\sigma\lambda} R_\nu^{\, \, \, \, \rho\sigma\lambda} - 2 R_{\mu\rho} R_\nu^{ \, \, \, \, \rho} - 2 R_{\mu\rho\nu\sigma} R^{\rho\sigma} + R R_{\mu\nu}.
\end{eqnarray}
It can be checked that background metric
\begin{eqnarray}
&& ds^2 = \frac{\ell^2}{z^2}\left[ - f(z) dt^2 + d\vec{x}^2 + \frac{dz^2}{f(z)}\right], \\  \label{GBgrav}
&& f(z) = \frac{1}{2\lambda_{\rm GB}\,f_0} \left[ 1 - \sqrt{1 - 4\lambda_{\rm GB} \left(1 -  z^4/z_h^4 \right)}\right],\nonumber \\
&&  f_0 \equiv\,\frac{1}{2\lambda_{\rm GB}} \left( 1 - \sqrt{1 - 4 \lambda_{\rm GB}} \right),  \quad \ell^2 \equiv \frac{L^2}{f_0} \ \label{fnot}
\end{eqnarray} obeys the equation of motion in (\ref{Eq:eomGB}) and that, in the $\lambda_{\rm GB} \to 0$ limit, we recover the AdS-Schwarzschild background. Additionally, in this geometry, the singularity is at infinity for positive $\lambda_{\rm GB}$, and at $z=\frac{z_h}{\sqrt{2} \lambda_{\rm GB}^{1/4}} (4\lambda_{\rm GB}-1)^{1/4}$.

The temperature in these coordinates is given by $1/T\,=\,\beta\,=\, \pi f_0 z_h$, and the entropy and energy densities are
\begin{equation}
s\,=\, \frac{\ell^3}{4G_N \, z_h^3},\qquad \mathcal{E}\,=\, \frac{3}{4} s\,T.
\end{equation}

Finally, the values of $\lambda_{\rm GB}$ were constrained by causality to the interval  \cite{Brigante:2008gz,Buchel:2009sk}
\begin{eqnarray}
- 7/36 \le \lambda_{\rm GB} \le 9/100.
\end{eqnarray}  However, more recently, it has been pointed out in \cite{Camanho:2014apa} that Gauss-Bonnet gravity, taken as an exact theory, violates causality for any value of $\lambda_{\rm GB}$.

\subsection{Mutual Information in Gauss-Bonnet gravity}\label{Sec:MI_NSW_HD}

Our main purpose in this section is to examine how mutual information is disrupted for the a two-sided black hole given by (\ref{GBgrav}). Following the prescription that we used for the non-conformal cases in Section (\ref{Sec:MI_NonSW_Dp}), we compute the mutual information between two intervals located in opposite sides of the black hole. We do this first for the unperturbed geometry and then for the metric that includes the backreaction of a shock wave coming from the left side of the hole.

The generalization of the Ryu-Takayanagi prescription to theories with Gauss-Bonnet gravity in the bulk has been proposed in \cite{deBoer:2011wk, Hung:2011xb}, building on \cite{Jacobson:1993xs}\footnote{The prescription for entanglement entropy in general theories with higher derivatives is given the Wald's entropy formula plus a term involving the extrinsic curvature \cite{Dong:2013qoa,Camps:2013zua}}. Given a boundary region $A$ and a three dimensional surface $\Sigma$, whose boundary coincides with the two dimensional boundary $\partial A$, the entanglement entropy $S_{\rm EE}(A)$ is now obtained by extremizing
\begin{eqnarray}
S_{\rm EE} = \frac{1}{4G_N} \int_\Sigma d^3 \sigma \sqrt{\tilde{\gamma}} \left( 1 + \lambda_{\rm GB}\,f_0 \ell^2 R_\Sigma \right) + \frac{\lambda_{\rm GB} \,f_0 \ell^2}{2 G_N} \int_{\partial \Sigma} d^2 \sigma \sqrt{h} K \ , \label{Eq:EEGB}
\end{eqnarray}
where $\sigma$ denotes the world-volume coordinates on $\Sigma$, and $\tilde{\gamma}$ is the induced metric on the surface. In addition, $R_\Sigma$ is the Ricci scalar corresponding to the induced metric on $\Sigma$ and the last term is the Gibbons-Hawking boundary term. \footnote{The holographic EE in a time dependent Gauss-Bonnet gravity background was studied recently in \cite{Caceres:2015bkr}.}

Let us first compute the entanglement entropy $S_A$ (or $S_B$) for the ``rectangular strip'' $A$, $z = z(x),\,x\equiv x_1 \in [-L/2,L/2]$, $x_{2,3}\,\in (-\infty,\,\infty)\,$. With this parametrization, the induced metric on the surface is given by
\begin{eqnarray}
ds_{\rm ind}^2 = \frac{\ell^2}{z^2} \left(dx_2^2 + dx_3^2 \right) + \frac{\ell^2}{z^2} \left(1 + \frac{z'^2}{f}\right) dx^2,\quad ' \equiv d/dx.
\end{eqnarray}

This yields the following effective area
\begin{eqnarray}
S_{\rm eff} = \frac{\ell^3\,\mathcal{V}}{4 G_N}   \int  \frac{dx}{z^3} \left[  \left(1+ \frac{z'^2}{f} \right)^{1/2} + 2 \lambda_{\rm GB} \frac{z'^2}{\left(1+ \frac{z'^2}{f} \right)^{1/2}} \right].  \label{Eq:OnshellSgb}
\end{eqnarray} Since $x$ is a cyclic coordinate, its conjugated momentum is conserved. Thus, the equation of motion that minimizes the area is
\begin{eqnarray}
\frac{\sqrt{f}}{z^3} \frac{ f + z'^2 - 2\lambda_{\rm GB} z'^2 f}{\left(f + z'^2\right)^{3/2}} = {\rm const} = \frac{1}{z_{\rm min}^3} \ ,
\end{eqnarray}
where we have imposed the boundary condition that at $z = z_{\rm min}$, $z'(x) \to 0$.  A more convenient way to parametrize the minimal surface is by using $x(z)$ instead. In such a choice, we obtain the momentum conservation equation
\begin{eqnarray}
\sqrt{f}\,x'(z) \left(\frac{1+ f( x'^2 - 2\lambda_{\rm GB})}{\left(1 +f\,x'^2\right)^{3/2}}\right) = \left( \frac{z}{z_{\rm min}}\right)^3 .
\end{eqnarray} By defining the rescaled coordinate $\tau\equiv z/z_*$ and $\zeta(\tau)\equiv x'^2$, we obtain the width of the intervals $A$ and $B$
\begin{equation}
L\,=\,2\int_0^1\,d\tau\,z_{\rm min}\,\sqrt{\zeta(\tau)},
\end{equation}  and the entanglement entropy is given by
\begin{equation}
S_{A}\,=\,S_{B}\,=\,\frac{\ell^3\,\mathcal{V}}{2G} \int^1_\epsilon d\tau\,\frac{z_{\rm min}^{-2}}{\tau^3}\left((1/f+\zeta)^{1/2} +\frac{2\lambda_{\rm GB}}{(1/f+\zeta)^{1/2}} \right).  \label{Eq:SA-HD}
\end{equation}

\begin{figure}
\begin{center}
 \includegraphics[scale=.35]{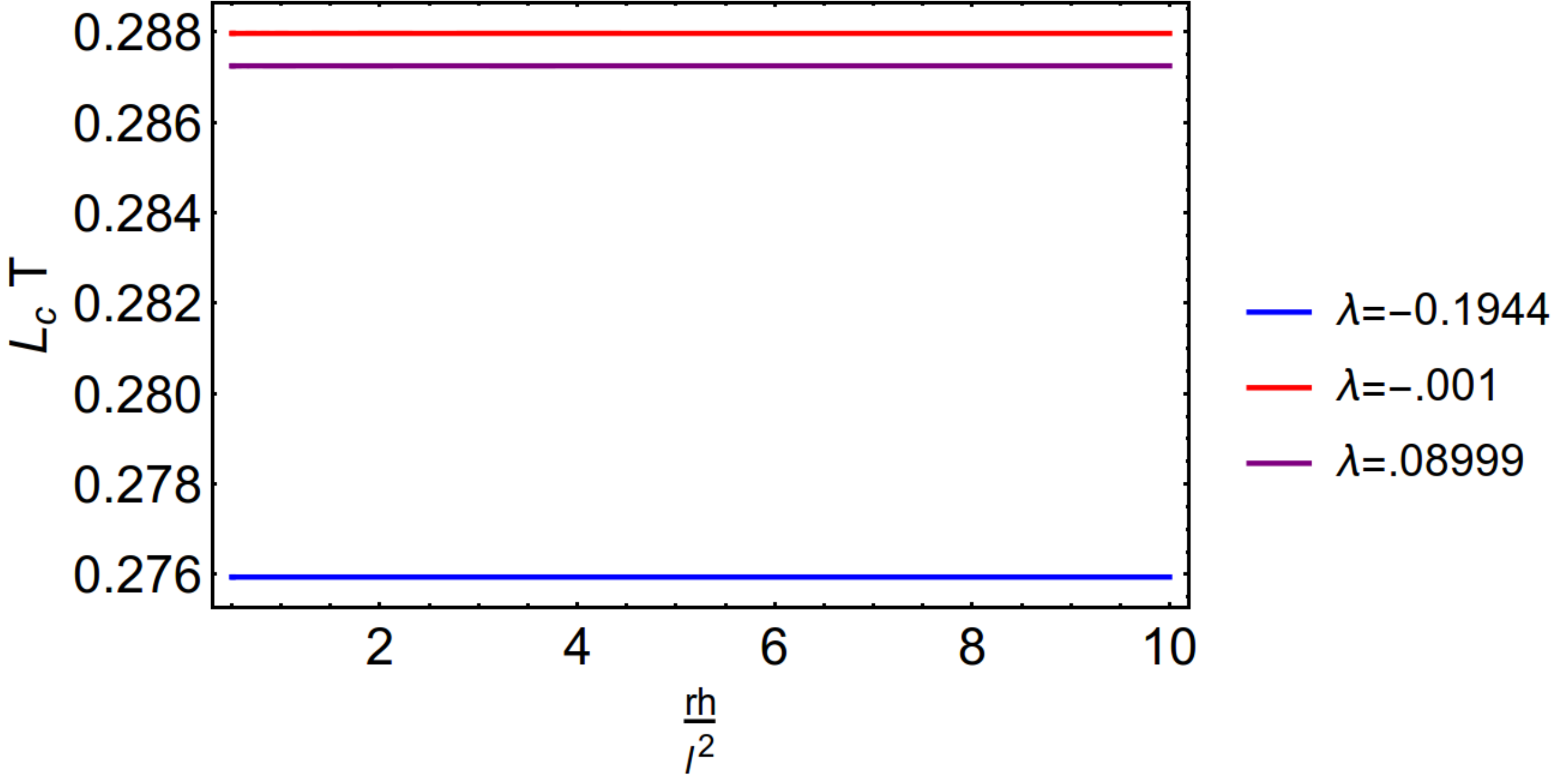}
 \caption{Critical values of $L$ as a function of the horizon $r_H\equiv \ell^2/z_h$, in units of $\ell$, for $\lambda_{\rm GB}\,=\,$ -0.1944 (blue), -0.0001(red), and 0.08999 (purple)}\label{Fig:LcritHD}
 \end{center}
\end{figure}

On the other hand, the entanglement entropy $S_{A \cup B}$ is given by four times the area of the surface that extends from $(z,\,x)\,=\,(0,\,L/2)$ (or equivalently $(0,\,0)$ due to translation symmetry) to the horizon, $(z,\,x)\,=\,(z_h,\,L/2)$. This results in
\begin{equation}
S_{A \cup B}\,=\,\frac{\ell^3\,\mathcal{V}}{G} \int^{z_h/z_{\rm min}}_\epsilon d\tau\,\frac{z_{\rm min}^{-2}}{\tau^3}\left(1/\sqrt{f} +2\lambda_{\rm GB} \sqrt{f} \right).
\end{equation}
Thus, the mutual information between the intervals is given by
\begin{equation}
I(A,B)\,=\,\frac{\ell^3\,\mathcal{V}}{G}  \left\lbrace  \begin{array}{l l}
 \int^1_\epsilon d\tau\,\frac{z_{\rm min}^{-2}}{\tau^3}\left((1/f+\zeta)^{1/2} +\frac{2\lambda_{\rm GB}}{(1/f+\zeta)^{1/2}} \right)-\int^{\frac{z_h}{z_{\rm min}}}_\epsilon d\tau\,\frac{z_{\rm min}^{-2}}{\tau^3}\left(1/\sqrt{f} +2\lambda_{\rm GB} \sqrt{f} \right), & {\rm if\, >\,0} \\
 0, \quad {\rm otherwise} &
 \end{array} \right. \label{Eq:MutInfoGB}
\end{equation} For low values of $z_{\rm min}$ ({\it i.e.} small width $L$), the mutual information vanishes since the sum of the individual entanglement entropies, $S_A$ and $S_B$, is smaller than $S_{A \cup B}$. However, as $z_{\rm min}$ approaches the horizon $z_h$, $I(A,B)$ becomes non-zero and is given by the first line in (\ref{Eq:MutInfoGB}). Figure \ref{Fig:LcritHD} shows the values of the interval width for which mutual information vanishes, for different values of $\lambda_{\rm GB}$.

\subsection{Shock wave geometry and scrambling time} \label{Sec:Scrambling-HD}

Let us now estimate the scrambling time by analyzing the perturbed geometry when a null shock wave is added from the left boundary (see Figure \ref{minimalsurf1}). First, for the sake of comparison with previous sections, let us define $r\,\equiv\,\ell^2/z$ and rewrite (\ref{GBgrav}) as
\begin{equation}
ds^2\,=\,-g(r) \,dt^2 \,+\,\frac{dr^2}{g(r)}\,+ \frac{r^2}{\ell^2} d\vec{x}^2,
\end{equation} where now
\begin{equation}
g(r)\equiv \frac{r^2}{\ell^2\,2\lambda_{\rm GB}\,f_0} \left(1-\sqrt{1-4\lambda_{\rm GB} \left(1- \frac{r_H^4}{r^4} \right)} \right).
\end{equation}

The temperature is given by
\begin{equation}
\beta\,\equiv\,\frac{4\pi}{g'(r_H)}\,=\,\frac{f_0\,\ell^2\,\pi}{r_H},
\end{equation} which depends on $\lambda_{\rm GB}$ through $f_0$ (see (\ref{fnot})).

The tortoise coordinate and the Kruskal coordinates are defined in the standard way. The backreaction of the shock wave into the geometry can be approximated again by a shift in the Kruskal time $v\rightarrow v+\alpha$. The tortoise coordinate $r_*$ is found to be
\begin{eqnarray}
 r_* &&\approx\,  \frac{f_0\,\ell^2}{4 r_H} \left( \log \left(\frac{r-r_H}{2 r_H}\right)\,- \pi/2 \, +\, 4\lambda_{\rm GB} \right) + \mathcal{O}(r-r_H)\\
&& \,\approx\,  \frac{\beta}{4\pi}\left( \log(r-r_H)+C_{\lambda_{\rm GB}}\right), \qquad C_{\lambda_{\rm GB}}\,=\, 4\lambda_{\rm GB}-\pi/2-\log 2 r_H.
\end{eqnarray} As in the previous section, the scrambling time is then estimated by setting $\alpha \approx \mathcal{O}(1)$, which leads to
\begin{equation}
t_*\,=\,\frac{\beta}{2\pi}\,\log\,\frac{e^{-C_{\lambda_{\rm GB}}}}{\delta r_H}\,\approx \, \frac{\beta}{2\pi} \log\,S_{BH} + \frac{\beta}{2\pi} \left(\frac{\pi}{2}-4\lambda_{\rm GB}  \right),
\end{equation} for small values of $|\lambda_{\rm GB}|$.

In the next subsection, we will compute numerically the evolution of mutual information as a function of $\alpha$ to confirm that, indeed, it vanishes as $\alpha$ approaches values of order $\mathcal{O}(1)$.

\subsection{Disruption of Mutual Information after the shock wave}\label{Sec:ExtremalSurface_HD}

Now we want to find the extremal surface that interpolates between the half plane on one side ($A_h$) to the other side ($B_h$) of the two-sided geometry. We use as (\ref{GBgrav}) as the bulk metric. Similar to previous sections, we parameterize the surface by $z(t)$, in which case the induced surface is
\begin{eqnarray}
ds_{\rm ind}^2 = \frac{\ell^2}{z(t)^2} \left(dx_2^2 + dx_3^2 \right) + \frac{\ell^2}{z(t)^2} \left(-f + \frac{\dot{z}^2}{f}\right) dt^2 \ ,
\end{eqnarray} and the effective action to be extremized is

\begin{equation}
S_{\rm eff} = \frac{\ell^3 \mathcal{V}}{4 G_N}   \int  \frac{dt}{z^3} \left[  \left(-f+ \frac{\dot{z}^2}{f} \right)^{1/2} + 2 \lambda_{\rm GB} \frac{\dot{z}^2}{\left(-f+ \frac{\dot{z}^2}{f} \right)^{1/2}} \right] \ .  \label{eebghalf}
\end{equation} Now the conserved quantity corresponding to the $t-$translation invariance is
\begin{equation}
\gamma\,\equiv\,\frac{ \sqrt{-f(z_0)}}{z_0^3}\,=\,\frac{f^2-\dot{z}^2(1-2\lambda_{\rm GB} \,f)}{z^3\left(-f+\frac{\dot{z}^2}{f}\right)^{3/2}},\qquad \dot{z}|_{z_0}\,=\,0.
\end{equation}

Using $t(z)$ to parameterize the surface, and defining $\eta(z)\equiv t'(z)^2$, we obtain
\begin{equation}
t(z)\,=\,\int_0^z\,d\tilde{z} \sqrt{\eta(\tilde{z})},
\end{equation} where $\eta(z)$ is obtained from the real solution to the equation
\begin{equation}
\gamma\, =\,\frac{(\eta\, f^2-(1-2\lambda_{\rm GB} \,f))\sqrt{\eta}}{z^3\left(-\eta\,f+\frac{1}{f}\right)^{3/2}}.
\end{equation}

In order to find a relation between $\alpha$ and $z_0$, we divide the extremal surface in three sections, in the same way as explained in Section \ref{Sec:Ext_Surf_HP_Dp}, using $dz_*\equiv -dz/f(z)$. In this case, we obtain

\begin{eqnarray}
&& \alpha\,=\,2\,e^{K_1+K_2+K_3} \\
&& K_1\,=\,-\frac{4\pi}{\beta}\int_{z_0}^{\bar{z}}\,dz\frac{1}{-f(z)}, \nonumber \\
&& K_2\,=\,\frac{2\pi}{\beta}\int_{0}^{z_h}\,dz\left(\frac{1}{-f(z)}-\sqrt{\eta(z)} \right), \nonumber \\
&& K_3\,=\,\frac{4\pi}{\beta}\int_{zh}^{z_0}\,dz\left(\frac{1}{-f(z)}-\sqrt{\eta(z)} \right), \nonumber
\end{eqnarray} where $\bar{z}$ is, again, a surface inside the black hole where $z_*=0$.

\begin{figure}[h]
\centering
\includegraphics[width=8.7cm]{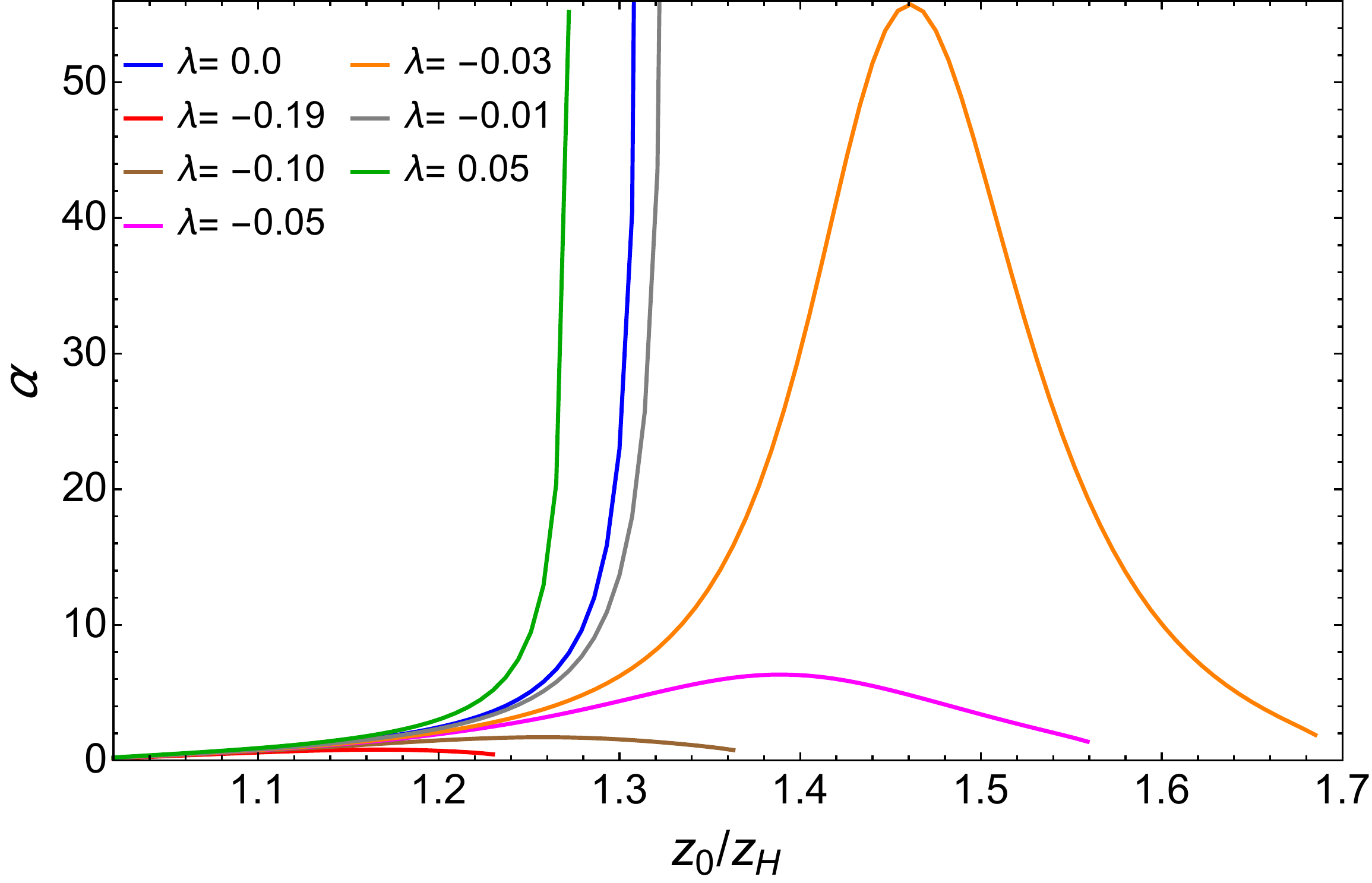}
\caption{Values of $\alpha$ as a function of $z_0$, for several values of $\lambda_{\rm GB}$. \label{Fig:alpha_HD}}
\end{figure}

Figure \ref{Fig:alpha_HD} depicts $\alpha$ as a function of the ``turning'' point $z_0$. In the case of positive $\lambda_{\rm GB}$ (green line), we see that the rapid growth is enhanced and qualitatively does not differ from the gravity calculations Einstein. However, a drastically different behavior is found for negative values of the Gauss-Bonnet coupling. In that case, $\alpha$ reaches a maximum value and then decreases until it reaches a critical value. This last value corresponds to the fact that $z_0$ has reached the singularity. We will see what this implies for EE and mutual information in the next paragraphs.

The entanglement entropy that corresponds to the extremal surface described above is given by
\begin{eqnarray}
S_{A_h \cup B_h}\,&=&\,\frac{\ell^3\,\mathcal{V}}{G_N}\left[ \int_{0}^{z_h}\,dz \frac{1}{z^3} \left(\left(-f \eta+1/f \right)^{1/2} +\frac{2\lambda_{\rm GB}}{\left(-f \eta+1/f \right)^{1/2}} \right)\right.  \label{Eq:SAB-HD}\\
&&\left. +2\int_{z_h}^{z_0}\,dz \frac{1}{z^3} \left(\left(-f \eta+1/f \right)^{1/2} +\frac{2\lambda_{\rm GB}}{\left(-f \eta+1/f \right)^{1/2}} \right)\right]. \nonumber
\end{eqnarray}

\begin{figure*}[h]
\centering
\includegraphics[width=8.5cm]{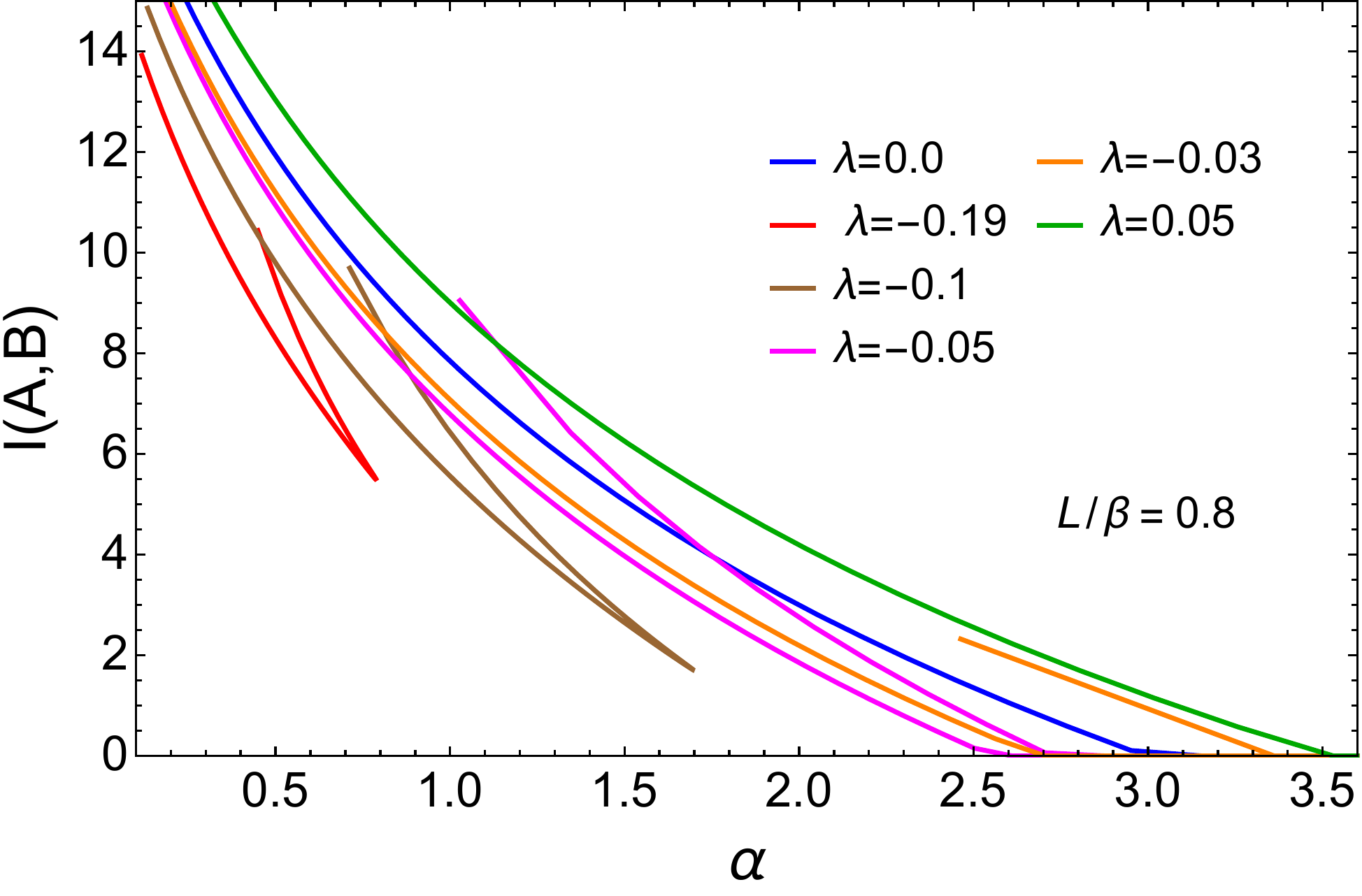}
\caption{Mutual information $I(A,B)$ as a function of $\alpha$  for several values of $\lambda_{\rm GB}$. In these plots, we used $L/\beta\,=\, 0.8$.\label{Fig:Sev_lambdas}}
\end{figure*}

\begin{figure*}[h]
\centering
\includegraphics[width=8.5cm]{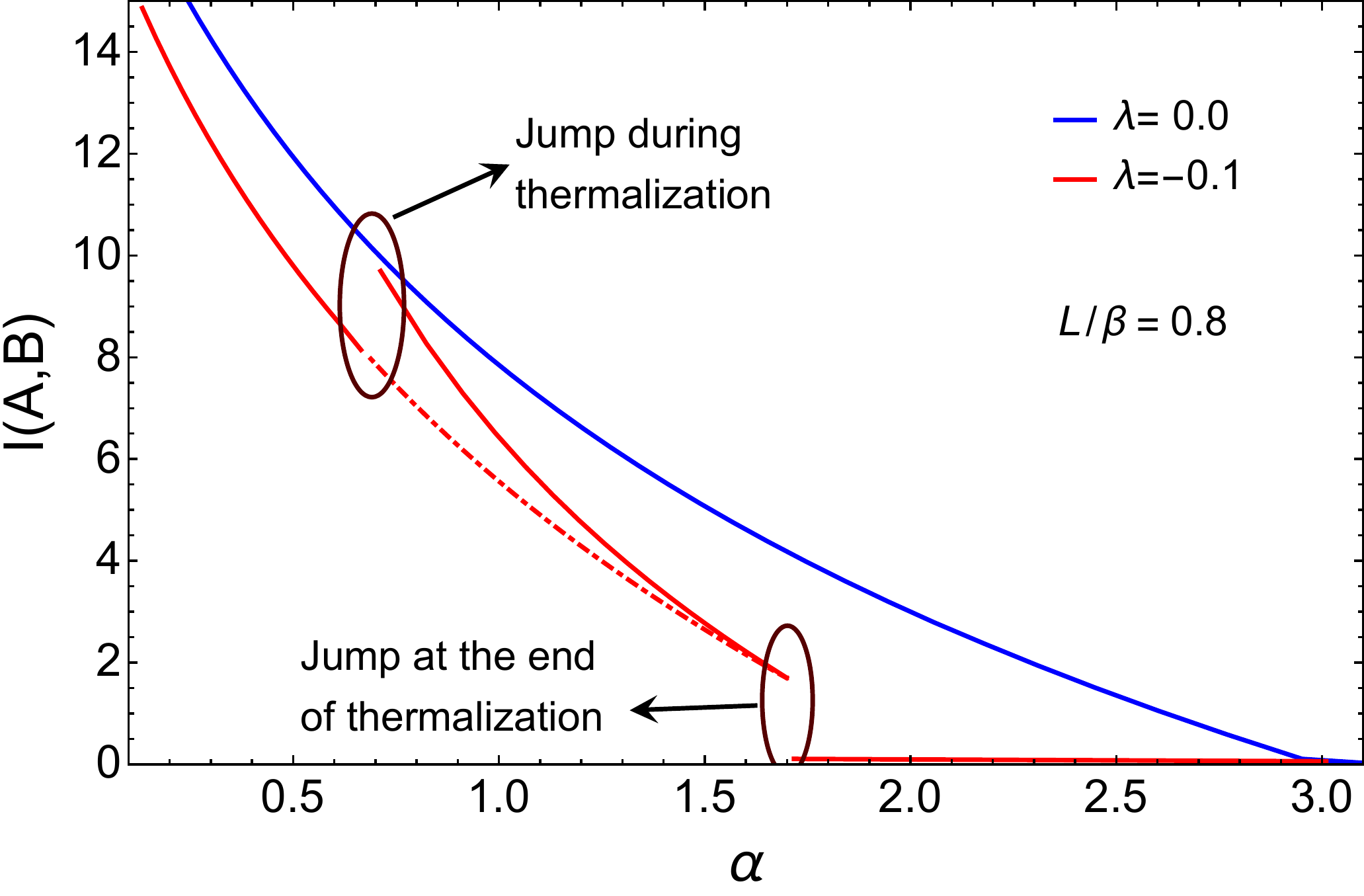}
\caption{Discontinuities in mutual information, $I(A,B)$, as a function of $\alpha$ for negative $\lambda_{\rm GB}$ (solid red line). The blue line shows the case of no Gauss-Bonnet coupling for the sake of comparison.\label{Fig:MI-jump}}
\end{figure*}

In Figure \ref{Fig:Sev_lambdas} we show mutual information as a function of $\alpha$, which is given now obtained by subtracting (\ref{Eq:SAB-HD}) from twice the result of (\ref{Eq:SA-HD}). For positive $\lambda_{\rm GB}$, the behavior is similar to the case of no Gauss-Bonnet coupling, and mutual information indeed vanishes for $\alpha\sim \mathcal{O}(1)$ as assumed in the calculation of the scrambling time in Section \ref{Sec:Scrambling-HD}. For negative $\lambda_{\rm GB}$, however, we notice here an anomalous behavior, due to the existence of two values of $z_0$ for a given value of $\alpha$ (see Figure \ref{Fig:alpha_HD}). Following the prescription from \cite{Hubeny:2007xt} to compute $S_{A_h \cup B_h}$, we must choose the surface with smaller area. This implies a ``jump'' in the evolution of entanglement entropy and mutual information as the dependence between the two sides is erased. Likewise, there is a second discontinuity at the end of scrambling which happens when the disconnected regions at each side of the eternal black black hole have a greater area than the surface that interpolates between the two boundaries. This is further illustrated in Figure \ref{Fig:MI-jump}. It is worth noticing that a similar behavior was also found in the analysis of EE in a Vaidya-like background in \cite{Caceres:2015bkr}.

\section{Outlook}~\label{Sec:summary}

In the  context of generalized holography (not restricted to asymptotically AdS space times), we have shown that the dual thermal field theory scrambles information
at time scales given by $\frac{\beta}{2\pi} \log S$. There are in general some sub-leading corrections which depend on the various parameters of the bulk geometry.
The scrambling behavior was studied (holographically) by following  time dependence of thermo-mutual information for some given region in the boundary theory. For
non-conformal and non-relativistic cases, the behavior was found to be similar to the conformal case with $a=c$. In case of theories dual to Gauss Bonnet Gravity, corresponding in particular
to the conformal case with $c \ne a$, the behavior is very different depending on the sign of Gauss Bonnet coupling. The results were summarized in the introduction to this article.
We would like to conclude by mentioning some of the open issues to which we want to return in future work.
\begin{itemize}
 \item Thermo-mutual information~\cite{Morrison:2012iz},  studied here as a probe of scrambling, involves regions in two different field theories of the thermofield
 double. It will be interesting to understand the map of thermo-mutual information to a probe defined only on one side, generalizing the map in case of $1+1$ dimensional
 CFT, studied in~\cite{Morrison:2012iz}.
 \item The scrambling time ($t_*$) is defined by considering vanishing of this thermo-mutual information~\cite{Shenker:2013pqa}. Alternatively scrambling time ($\tilde t_*$)
 is also defined by exponential fall of a function $F(t)$~\cite{Maldacena:2015waa} defined in terms of four point functions in the field theory. In  the case of the
 BTZ black hole and its dual theory, it can be shown~\cite{Shenker:2013pqa} that the two scrambling times are essentially  the same, at least in case of some heavy operators. However, a precise
 connection between the two is still missing in the literature.
 \item In the alternate probe of scrambling in terms of four point functions, the Lyapunov exponent provides the rate of exponential decrease of $F(t)$ at early times \cite{Maldacena:2015waa}; whereas the late-time decay is characterized by the poles of the Green's function, which holographically correspond to quasinormal modes. In \cite{Maldacena:2015waa}, a bound on the Lyapunov exponent was also proposed. A precise definition of the Lyapunov exponent
 in terms of the behavior of mutual information would be very useful.
 \item In this note ,we considered the Gauss Bonnet higher curvature gravitational theory. An interesting question is obviously how would the mutual information behave in the cases of other higher curvature corrections to the gravitational action.

\end{itemize}




\section*{Acknowledgments}~\label{Sec:ackn}

We would like to thank specially to S. Yankielowicz for very productive and encouraging discussions. We are grateful to O. Aharony, O. Ben-Ami, E. C\'{a}ceres, D. Carmi, G. Gur-Ari, D. Harlow, N. Itzhaki, V. Kaplunovsky, L. Pando Zayas, and V. Rosenhaus, for insightful conversations and comments.
This work was supported in part by a centre of excellence supported by the Israel Science Foundation (grant number 1989/14),
and by the US-Israel bi-national fund (BSF) grant number 2012383 and the Germany Israel
bi-national fund GIF grant number I-244-303.7-2013. The work of N.S was supported by "The PBC program for fellowships for outstanding post-doctoral researcher from China and
India of the Israel council of higher education''.



\appendix

\section{A series expansion of EE in $Dp$-brane backgrounds}\label{Appendix:SeriesExp-Dp}

In this appendix we include a series approximation of the entanglement entropy of one strip, as a supplement to the numerical calculation of Section \ref{Sec:MI_NonSW_Dp}. There, we obtained that  $S_A$ or $S_B$, with width $L$, is given by the extremized area
\begin{equation}
{\rm Area}_{d-2}\,=\,2\frac{\mathcal{V}}{\ell^{d-1}} \ell^{8-p-\sigma} \int_{r_{\rm min}}^{\infty}dr \frac{r^\sigma}{\sqrt{1-\frac{r_H^s}{r^s}}} \frac{1}{\sqrt{1-\left( \frac{r_{\rm min}}{r}\right)^{\omega}}}, \label{Eq:area_Einst_App}
\end{equation}
where \begin{eqnarray}
&& \sigma\, \equiv\, \frac{n}{16}(7-p)^2(p-2)+ \frac{n}{16}(p-3)^2(8-p) ,\quad s\,\equiv \, n(7-p), \label{Eq:def_parameters1_App}\\
&& \omega\,\equiv\, \frac{p\,n}{8}(7-p)^2+ \frac{n}{8}(p-3)^2(8-p),  \nonumber
\end{eqnarray}
and $r=\Lambda$ is set as the cutoff. The divergent part of the area yields
\begin{equation}
{\rm Area}_{\rm div} \,=\,2\frac{\mathcal{V}}{\ell^{d-1}} \ell^{8-p-\sigma} \frac{\Lambda^{\sigma+1}}{\sigma+1},
\end{equation}
and the width $L$ is given by
\begin{equation}
L\,=\,2\int_{r_{\rm min}}^{\infty} dr \, \left(\frac{r}{\ell}\right)^{-2\xi}  \frac{1}{\sqrt{1-\frac{r_H^s}{r^s}} \sqrt{\left( \frac{r}{r_{\rm min}}\right)^{\omega} -1 }}.  \label{Eq:L_Einst_App}
\end{equation}

Using the coordinate redefinition $u\,\equiv \, \frac{r_{\rm min}}{r}$, we rewrite the area as
\begin{equation}
{\rm Area}^{\rm finite}_{d-2}\,=\,2\frac{\mathcal{V}}{\ell^{d-1}} \ell^{8-p-\sigma} r_{\rm min}^{1+\sigma}  \int_{r_{\rm min}/\Lambda}^{1} du \frac{u^{-2-\sigma}}{\sqrt{1-u^\omega}} \frac{1}{\sqrt{1-\left( \frac{r_H}{r_{\rm min}}u\right)^{s}}} - {\rm Area}_{\rm div}, \label{Eq:area_u_App}
\end{equation} and the lengths as
\begin{equation}
L\,=\,2 \ell^{2\xi}r_{\rm min}^{1-2\xi}\int_{0}^{1} du \, u^{-2+4\gamma+\sigma}  \frac{1}{\sqrt{ 1 - u^{\omega} } \sqrt{1- \frac{r_H^s}{r_{\rm min}^s}u^s} }.  \label{Eq:L_u_App}
\end{equation}

Now we would like to have some analytic approximation for these results. To do so, we use the fact that $r_{\rm min} > r_H$, and perform the series expansion \cite{Fischler:2012ca}
\be
 \frac{1}{\sqrt{1-\left( \frac{r_H}{r_{\rm min}}u\right)^s}} \,=\, \sum_{n=0}^\infty \frac{\Gamma [n+1/2] }{\sqrt{\pi} \Gamma [n+1]} \left( \frac{r_H}{r_{\rm min}}\right)^{n\,s} u^{n\,s},
\ee which results in
\begin{eqnarray}
{\rm Area}^{\rm finite}_{d-2}\,=&& 2\frac{\mathcal{V}}{\ell^{d-1}} \ell^{8-p-\sigma} r_{\rm min}^{1+\sigma}   \left[
-\frac{\sqrt{\pi} \,\Gamma \left[1-\frac{1+\sigma}{\omega}\right] }{(1+\sigma)\Gamma \left[\frac{1}{2}-\frac{1+\sigma}{\omega}\right]} \label{Eq:Area_sum} \right. \\
&& \left.+\,\sum_{n=1}^\infty \frac{\Gamma [n+1/2]  \Gamma \left[\frac{n\,s\,-1-\sigma}{\omega}\right]}{ \omega\,\Gamma [n+1]  \Gamma \left[\frac{2\,n\,s\,-2-2\sigma\,+\,\omega}{2\,\omega}\right]} \left( \frac{r_H}{r_{\rm min}}\right)^{n\,s} \right], \nonumber
\end{eqnarray} and

\begin{equation}
L\,=\,2 \ell^{2\xi}r_{\rm min}^{1-2\xi}\, \sum_{n=0}^\infty \frac{\Gamma [n+1/2]  \Gamma \left[\frac{n\,s\,-1+4\xi - \sigma}{\omega}\right]}{\omega\, \Gamma [n+1]  \Gamma \left[\frac{2\,n\,s\,-2+2(4\xi-\sigma)\,+\,\omega}{2\,\omega}\right]} \left( \frac{r_H}{r_{\rm min}}\right)^{n\,s} .
\end{equation}

Both expressions are convergent and finite for the case $r_{\rm min} \gg r_H$. But we are interested in the case where  $r_{\rm min} \sim r_H$, since this is the regime where mutual information will become very small. In order to obtain a convergent expression, we need to reorganize (\ref{Eq:Area_sum}).  Using the properties of the gamma function we can rewrite the coefficients of the sum in (\ref{Eq:Area_sum}) and write the area as

\begin{eqnarray}
{\rm Area}^{\rm finite}_{d-2}\,=&& 2\frac{\mathcal{V}}{\ell^{d-1}} \ell^{8-p-\sigma} r_{\rm min}^{1+\sigma}   \left[ \left(\frac{r_{\rm min}}{\ell} \right)^{2\xi} \frac{L}{2\,r_{\rm min}}
-\frac{\sqrt{\pi}\,(\omega-2\sigma) \,\Gamma \left[1-\frac{1+\sigma}{\omega}\right] }{\omega \,\Gamma \left[\frac{3\omega\,-\,2(\sigma+1)}{2\omega}\right]} \label{Eq:Area_sum2} \right. \\
&& \left.+\,\sum_{n=1}^\infty \frac{1}{n\,s\,+\,(\sigma +1)} \frac{\Gamma [n+1/2]  \Gamma \left[\frac{n\,s\,-1-\sigma}{\omega}\right]}{ \Gamma [n+1]  \Gamma \left[\frac{2\,n\,s\,-2-2\sigma\,+\,\omega}{2\,\omega}\right]} \left( \frac{r_H}{r_{\rm min}}\right)^{n\,s} \right], \nonumber
\end{eqnarray} which is now convergent for in the limit $r_{\rm min} \sim r_H$. In such a limit, we obtain

\begin{eqnarray}
&& {\rm Area}^{\rm finite}_{d-2}\,=\, 2\frac{\mathcal{V}}{\ell^{d-1}} \ell^{8-p-\sigma} r_H^{1+\sigma}   \left( \left(\frac{r_H}{\ell} \right)^{2\gamma} \frac{L}{2\,r_H} \,+\, \mathcal{A}_1 \label{Eq:Area_large} \right), \\
&& \mathcal{A}_1 \,\equiv \,-\frac{\sqrt{\pi}\,(\omega-2\sigma) \,\Gamma \left[1-\frac{1+\sigma}{\omega}\right] }{\omega \,\Gamma \left[\frac{3\omega\,-\,2(\sigma+1)}{2\omega}\right]}\,+\,\sum_{n=1}^\infty \frac{1}{n\,s\,+\,(\sigma +1)} \frac{\Gamma [n+1/2]  \Gamma \left[\frac{n\,s\,-1-\sigma}{\omega}\right]}{ \Gamma [n+1]  \Gamma \left[\frac{2\,n\,s\,-2-2\sigma\,+\,\omega}{2\,\omega}\right]}. \nonumber
\end{eqnarray}

Now, we compute the area of the surface that interpolates between the two boundaries. This area corresponds to four times the area of a surface that divides the boundary in half and extends from one boundary to the black hole. Defining $\tilde{u} \equiv \,u/r_H$,

\begin{eqnarray}
{\rm Area}_{A \cup B}=&&4\,\frac{\mathcal{V}}{\ell^{d-1}} \ell^{8-p-\sigma} r_H^{\sigma+1}\int_{r_H/\Lambda}^{1} d\tilde{u} \frac{\tilde{u}^{-2-\sigma}}{\sqrt{1-\tilde{u}^s}}  - 2\,{\rm Area}_{\rm div}, \label{Eq:area_AuB_App} \\
=&& 4\,\frac{\mathcal{V}}{\ell^{d-1}} \ell^{8-p-\sigma} r_H^{\sigma+1} \frac{\sqrt{\pi } \Gamma \left[-\frac{\sigma +1}{s}\right]}{s \Gamma \left[\frac{s-2 \sigma -2}{2 s}\right]}\,\,\,\equiv\,\, 4\,\frac{\mathcal{V}}{\ell^{d-1}} \ell^{8-p-\sigma} r_H^{\sigma+1} \,\mathcal{A}_2
\end{eqnarray}

Thus, the mutual information between $A$ and $B$ is given by
\be
I(A,B)\,\equiv\, S_A+S_B-S_{A\cup B}\,=\, \frac{\mathcal{V}}{\ell^{d-1}G_N} \ell^{8-p-\sigma} r_H^{\sigma+1} \left(  \left(\frac{r_H}{\ell} \right)^{2\xi} \frac{L}{2\,r_H} \,+\, \mathcal{A}_1\,-\,\mathcal{A}_2 \right).
\ee

\section{Two sided entanglement entropy for half plane in the case of $D5$ brane}\label{Appendix:AnalyticExtSurf-D5}
Let us consider the case with $p=5$, \emph{i.e.} $D5$ brane. We can repeat the calculations in Sec.(\ref{Sec:Ext_Surf_HP_Dp}) in this case, but now analytically.
We have used these results to do a consistancy check on our numerics.
Let us define $\tilde \gamma^2 =\frac{2 R \gamma^2}{r_0^8}=\frac{r_h^4}{r_0^4}-1\ge0$ (as $r_0<r_h$). The inverse temperature
is given by $\beta= 4\pi \ell$ (note that it does not depend on the horizon radius). We have used this analytical results as a reference for our numerical calculations.
We can get an analytic solution for minimal surface as,
\be
t(x=\frac{r}{r_h})=\ell \log(|x^4-1|)-\ell \log\left| x^4-1+2 \tilde \gamma^2+\tilde \gamma \sqrt{4 \tilde\gamma^2+x^4(x^4-1)}\right| .
\ee
Also,
\be
r_*(x=\frac{r}{r_h})=\int \frac{dr}{f(r)}=\ell \log(|x^4-1|) .
\ee
We will choose $\bar x=\frac{\bar r}{r_h}=0$ in the following calculations.
The various integrals ($x_0=\frac{r_0}{r_h}$, $x_{\Lambda}=\frac{r_{\Lambda}}{r_h}$),
\bea
K_1 &=& \log(1-x_0^4),\\
K_2 &\simeq& 2 \log x_{\Lambda}+ \half \log\left( \frac{1+2 x_0^2 \sqrt{1-x_0^4}}{4 x_0^4 (1-x_0^4)}\right),\,~~ x_{\Lambda} \gg 1 ,\\
K_3 &=& \log \left( \frac{4x_0^4}{2 x_0^4-1} \right),\\
u_{\Lambda} &\simeq& x_{\Lambda}^2,\,~~ x_{\Lambda} \gg 1.
\eea
Then,
\be
\alpha =\frac{4  x_0^2 \sqrt{1-x_0^4} \sqrt{1+2 x_0^2 \sqrt{1-x_0^4}}}{2 x_0^4-1}.
\ee
This equations clearly shows that there is critical value of $x_0=x_{\text{crit}}=2^{-\frac{1}{4}}$ beyond which surface does
not penetrate the horizon.

We can invert this relation to get $x_0$ in terms of $\alpha$,
\be
x_0^2=\frac{1}{8} \left( -\alpha + \sqrt{16+\alpha^2} +\sqrt{16 -2 \alpha^2+ 2\alpha \sqrt{16+\alpha^2}}\right).
\ee
We can now calculate the area for this surface which will correspond to $S_{A \cup B}$. This has two ultraviolet $\alpha$
independent divergence ($\sim \Lambda^4, ~~\log(\Lambda)$) piece.
The regularized entanglement entropy (defined in \eqref{Eq:RegEEHP-Dp}) is given by,
\be
S_{A_h \cup B_h}^{reg}(\alpha)=- n_5 \left[ \half \alpha (\alpha -\sqrt{16+\alpha^2})-8 \log\left(\alpha +\sqrt{16+\alpha^2}\right)+16 \log2\right],
\ee
where $n_5=\frac{1}{64} \frac{\mathcal{V} r_h^4}{4 G_{N}}$. The regularized entanglement entropy  for large $\alpha$ grows as $\log \alpha$.





\begin{thebibliography}{9}

\bibitem{Ryu:2006bv}
  S.~Ryu and T.~Takayanagi,
  ``Holographic derivation of entanglement entropy from AdS/CFT,''
  Phys.\ Rev.\ Lett.\  {\bf 96}, 181602 (2006)
  [hep-th/0603001].

\bibitem{Ryu:2006ef}
  S.~Ryu and T.~Takayanagi,
  ``Aspects of Holographic Entanglement Entropy,''
  JHEP {\bf 0608}, 045 (2006)
  [hep-th/0605073].

\bibitem{Hubeny:2007xt}
  V.~E.~Hubeny, M.~Rangamani and T.~Takayanagi,
  ``A Covariant holographic entanglement entropy proposal,''
  JHEP {\bf 0707}, 062 (2007)
  [arXiv:0705.0016 [hep-th]].

\bibitem{Horowitz:1999jd}
  G.~T.~Horowitz and V.~E.~Hubeny,
  ``Quasinormal modes of AdS black holes and the approach to thermal equilibrium,''
  Phys.\ Rev.\ D {\bf 62}, 024027 (2000)
  [hep-th/9909056].

\bibitem{Balasubramanian:2011ur}
  V.~Balasubramanian {\it et al.},
  ``Holographic Thermalization,''
  Phys.\ Rev.\ D {\bf 84}, 026010 (2011)
  [arXiv:1103.2683 [hep-th]].

\bibitem{Allais:2011ys}
  A.~Allais and E.~Tonni,
  ``Holographic evolution of the mutual information,''
  JHEP {\bf 1201}, 102 (2012)
  [arXiv:1110.1607 [hep-th]].

\bibitem{Callan:2012ip}
  R.~Callan, J.~Y.~He and M.~Headrick,
  ``Strong subadditivity and the covariant holographic entanglement entropy formula,''
  JHEP {\bf 1206}, 081 (2012)
  [arXiv:1204.2309 [hep-th]].

\bibitem{Galante:2012pv}
  D.~Galante and M.~Schvellinger,
  ``Thermalization with a chemical potential from AdS spaces,''
  JHEP {\bf 1207}, 096 (2012)
  [arXiv:1205.1548 [hep-th]].

\bibitem{Caceres:2012em}
  E.~Caceres and A.~Kundu,
  ``Holographic Thermalization with Chemical Potential,''
  JHEP {\bf 1209}, 055 (2012)
  [arXiv:1205.2354 [hep-th]].

\bibitem{Caceres:2013dma}
  E.~Caceres, A.~Kundu, J.~F.~Pedraza and W.~Tangarife,
  ``Strong Subadditivity, Null Energy Condition and Charged Black Holes,''
  JHEP {\bf 1401}, 084 (2014)
  [arXiv:1304.3398 [hep-th]].


\bibitem{Liu:2013iza}
  H.~Liu and S.~J.~Suh,
  ``Entanglement Tsunami: Universal Scaling in Holographic Thermalization,''
  Phys.\ Rev.\ Lett.\  {\bf 112}, 011601 (2014)
  [arXiv:1305.7244 [hep-th]].

\bibitem{Liu:2013qca}
  H.~Liu and S.~J.~Suh,
  ``Entanglement growth during thermalization in holographic systems,''
  Phys.\ Rev.\ D {\bf 89}, no. 6, 066012 (2014)
  [arXiv:1311.1200 [hep-th]].


\cite{Kundu:2016cgh}
\bibitem{Kundu:2016cgh}
  S.~Kundu and J.~F.~Pedraza,
  ``Spread of entanglement for small subsystems in holographic CFTs,''
  arXiv:1602.05934 [hep-th].


\bibitem{Hayden:2007cs}
  P.~Hayden and J.~Preskill,
  ``Black holes as mirrors: Quantum information in random subsystems,''
  JHEP {\bf 0709}, 120 (2007)
  [arXiv:0708.4025 [hep-th]].

\bibitem{Sekino:2008he}
  Y.~Sekino and L.~Susskind,
  JHEP {\bf 0810}, 065 (2008)
  [arXiv:0808.2096 [hep-th]].

\bibitem{Shenker:2013pqa}
  S.~H.~Shenker and D.~Stanford,
  ``Black holes and the butterfly effect,''
  JHEP {\bf 1403}, 067 (2014)
  [arXiv:1306.0622 [hep-th]].

\bibitem{Maldacena:2001kr}
  J.~M.~Maldacena,
  ``Eternal black holes in anti-de Sitter,''
  JHEP {\bf 0304}, 021 (2003)
    [hep-th/0106112].

\bibitem{Morrison:2012iz}
  I.~A.~Morrison and M.~M.~Roberts,
  ``Mutual information between thermo-field doubles and disconnected holographic boundaries,''
  JHEP {\bf 1307}, 081 (2013)
  doi:10.1007/JHEP07(2013)081
  [arXiv:1211.2887 [hep-th]].

\bibitem{Swinney}
A. Fraser and H. Swinney,
``Independent coordinates for strange attractors from mutual information'',
  Phys.\ Rev.\ A {\bf 33}, 1134 (1986).


\bibitem{Caputa:2015waa}
  P.~Caputa, J.~Sim\'{o}n, A.~\v{S}tikonas, T.~Takayanagi and K.~Watanabe,
  ``Scrambling time from local perturbations of the eternal BTZ black hole,''
  JHEP {\bf 1508}, 011 (2015)
  [arXiv:1503.08161 [hep-th]].
  
\bibitem{Hosur:2015ylk} 
  P.~Hosur, X.~L.~Qi, D.~A.~Roberts and B.~Yoshida,
  ``Chaos in quantum channels,''
  JHEP {\bf 1602}, 004 (2016)
  [arXiv:1511.04021 [hep-th]].

\bibitem{Leichenauer:2014nxa}
  S.~Leichenauer,
  ``Disrupting Entanglement of Black Holes,''
  Phys.\ Rev.\ D {\bf 90}, no. 4, 046009 (2014)
  [arXiv:1405.7365 [hep-th]].

\bibitem{Shenker:2013yza}
  S.~H.~Shenker and D.~Stanford,
  ``Multiple Shocks,''
  JHEP {\bf 1412}, 046 (2014)
  [arXiv:1312.3296 [hep-th]].

\bibitem{Roberts:2014isa}
  D.~A.~Roberts, D.~Stanford and L.~Susskind,
  ``Localized shocks,''
  JHEP {\bf 1503}, 051 (2015)
  [arXiv:1409.8180 [hep-th]].


\bibitem{Shenker:2014cwa}
  S.~H.~Shenker and D.~Stanford,
  ``Stringy effects in scrambling,''
  JHEP {\bf 1505}, 132 (2015)
  [arXiv:1412.6087 [hep-th]].

\bibitem{Maldacena:2015waa}
  J.~Maldacena, S.~H.~Shenker and D.~Stanford,
  ``A bound on chaos,''
  arXiv:1503.01409 [hep-th].

\bibitem{KalyanaRama:1999zj} 
  S.~Kalyana Rama and B.~Sathiapalan,
  ``On the role of chaos in the AdS / CFT connection,''
  Mod.\ Phys.\ Lett.\ A {\bf 14}, 2635 (1999)
  [hep-th/9905219].


\bibitem{Itzhaki:1998dd}
  N.~Itzhaki, J.~M.~Maldacena, J.~Sonnenschein and S.~Yankielowicz,
  ``Supergravity and the large N limit of theories with sixteen supercharges,''
  Phys.\ Rev.\ D {\bf 58}, 046004 (1998)
  [hep-th/9802042].

\bibitem{Balasubramanian:2009rx}
  K.~Balasubramanian and J.~McGreevy,
  ``An Analytic Lifshitz black hole,''
  Phys.\ Rev.\ D {\bf 80}, 104039 (2009)
  [arXiv:0909.0263 [hep-th]].

  \bibitem{Cai:2001dz}
  R.~G.~Cai,
  ``Gauss-Bonnet black holes in AdS spaces,''
  Phys.\ Rev.\ D {\bf 65}, 084014 (2002)
  [hep-th/0109133].

\bibitem{Myers:2010ru}
  R.~C.~Myers and B.~Robinson,
  ``Black Holes in Quasi-topological Gravity,''
  JHEP {\bf 1008}, 067 (2010)
  [arXiv:1003.5357 [gr-qc]].


\bibitem{deBoer:2011wk}
  J.~de Boer, M.~Kulaxizi and A.~Parnachev,
  ``Holographic Entanglement Entropy in Lovelock Gravities,''
  JHEP {\bf 1107}, 109 (2011)
  [arXiv:1101.5781 [hep-th]].

\bibitem{Hung:2011xb}
  L.~-Y.~Hung, R.~C.~Myers and M.~Smolkin,
  ``On Holographic Entanglement Entropy and Higher Curvature Gravity,''
  JHEP {\bf 1104}, 025 (2011)
  [arXiv:1101.5813 [hep-th]].

\bibitem{Myers:2012ed}
  R.~C.~Myers and A.~Singh,
  ``Comments on Holographic Entanglement Entropy and RG Flows,''
  JHEP {\bf 1204}, 122 (2012)
  [arXiv:1202.2068 [hep-th]].

\bibitem{Barbon:2008sr}
  J.~L.~F.~Barbon and C.~A.~Fuertes,
  ``A Note on the extensivity of the holographic entanglement entropy,''
  JHEP {\bf 0805}, 053 (2008)
  [arXiv:0801.2153 [hep-th]].

\bibitem{Kol:2014nqa}
  U.~Kol, C.~Nunez, D.~Schofield, J.~Sonnenschein and M.~Warschawski,
  ``Confinement, Phase Transitions and non-Locality in the Entanglement Entropy,''
  JHEP {\bf 1406}, 005 (2014)
  [arXiv:1403.2721 [hep-th]].

\bibitem{Brigante:2008gz}
  M.~Brigante, H.~Liu, R.~C.~Myers, S.~Shenker and S.~Yaida,
  ``The Viscosity Bound and Causality Violation,''
  Phys.\ Rev.\ Lett.\  {\bf 100}, 191601 (2008)
  [arXiv:0802.3318 [hep-th]].

\bibitem{Caceres:2015bkr}
  E.~Caceres, M.~Sanchez and J.~Virrueta,
  ``Holographic Entanglement Entropy in Time Dependent Gauss-Bonnet Gravity,''
  arXiv:1512.05666 [hep-th].

\bibitem{Camanho:2014apa}
  X.~O.~Camanho, J.~D.~Edelstein, J.~Maldacena and A.~Zhiboedov,
  ``Causality Constraints on Corrections to the Graviton Three-Point Coupling,''
  JHEP {\bf 1602}, 020 (2016)
  [arXiv:1407.5597 [hep-th]].









\bibitem{Klebanov:2007ws}
  I.~R.~Klebanov, D.~Kutasov and A.~Murugan,
  ``Entanglement as a probe of confinement,''
  Nucl.\ Phys.\ B {\bf 796}, 274 (2008)
  [arXiv:0709.2140 [hep-th]].

\bibitem{Faraggi:2007fu}
  I.~Bah, A.~Faraggi, L.~A.~Pando Zayas and C.~A.~Terrero-Escalante,
  ``Holographic entanglement entropy and phase transitions at finite temperature,''
  Int.\ J.\ Mod.\ Phys.\ A {\bf 24}, 2703 (2009)
  [arXiv:0710.5483 [hep-th]].


\bibitem{vanNiekerk:2011yi}
  A.~van Niekerk,
  ``Entanglement Entropy in NonConformal Holographic Theories,''
  arXiv:1108.2294 [hep-th].

\bibitem{Kachru:2008yh}
  S.~Kachru, X.~Liu and M.~Mulligan,
  ``Gravity duals of Lifshitz-like fixed points,''
  Phys.\ Rev.\ D {\bf 78}, 106005 (2008)
  [arXiv:0808.1725 [hep-th]].


\bibitem{Keranen:2011xs}
  V.~Keranen, E.~Keski-Vakkuri and L.~Thorlacius,
  ``Thermalization and entanglement following a non-relativistic holographic quench,''
  Phys.\ Rev.\ D {\bf 85}, 026005 (2012)
  [arXiv:1110.5035 [hep-th]].



\bibitem{Boonstra:1998mp}
  H.~J.~Boonstra, K.~Skenderis and P.~K.~Townsend,
  ``The domain wall / QFT correspondence,''
  JHEP {\bf 9901}, 003 (1999)
  [hep-th/9807137].


\bibitem{Fischler:2012ca}
  W.~Fischler and S.~Kundu,
  ``Strongly Coupled Gauge Theories: High and Low Temperature Behavior of Non-local Observables,''
  JHEP {\bf 1305}, 098 (2013)
  [arXiv:1212.2643 [hep-th]].

\bibitem{'tHooft:1996tq}
  G.~'t Hooft,
  ``The Scattering matrix approach for the quantum black hole: An Overview,''
  Int.\ J.\ Mod.\ Phys.\ A {\bf 11}, 4623 (1996)
  [gr-qc/9607022].

\bibitem{Itzhaki:1996rb}
  N.~Itzhaki,
  ``Some remarks on 't Hooft's S matrix for black holes,''
  hep-th/9603067.



\bibitem{Jacobson:1993xs}
  T.~Jacobson and R.~C.~Myers,
  ``Black hole entropy and higher curvature interactions,''
  Phys.\ Rev.\ Lett.\  {\bf 70}, 3684 (1993)
  [hep-th/9305016].

\bibitem{Dong:2013qoa}
  X.~Dong,
  ``Holographic Entanglement Entropy for General Higher Derivative Gravity,''
  JHEP {\bf 1401}, 044 (2014)
  [arXiv:1310.5713 [hep-th]].


\bibitem{Camps:2013zua}
  J.~Camps,
  ``Generalized entropy and higher derivative Gravity,''
  JHEP {\bf 1403}, 070 (2014)
  [arXiv:1310.6659 [hep-th]].



\bibitem{Buchel:2009sk}
  A.~Buchel, J.~Escobedo, R.~C.~Myers, M.~F.~Paulos, A.~Sinha and M.~Smolkin,
  ``Holographic GB gravity in arbitrary dimensions,''
  JHEP {\bf 1003}, 111 (2010)
  [arXiv:0911.4257 [hep-th]].





\bibitem{Taylor:2008tg}
  M.~Taylor,
  ``Non-relativistic holography,''
  arXiv:0812.0530 [hep-th].

\bibitem{Taylor:2015glc}
  M.~Taylor,
  ``Lifshitz holography,''
  Class.\ Quant.\ Grav.\  {\bf 33}, no. 3, 033001 (2016)
  [arXiv:1512.03554 [hep-th]].


\bibitem{Alishahiha:2014cwa}
  M.~Alishahiha, A.~F.~Astaneh and M.~R.~M.~Mozaffar,
  ``Thermalization in backgrounds with hyperscaling violating factor,''
  Phys.\ Rev.\ D {\bf 90}, no. 4, 046004 (2014)
  [arXiv:1401.2807 [hep-th]].


\bibitem{Fonda:2014ula}
  P.~Fonda, L.~Franti, V.~Ker\"{a}nen, E.~Keski-Vakkuri, L.~Thorlacius and E.~Tonni,
  ``Holographic thermalization with Lifshitz scaling and hyperscaling violation,''
  JHEP {\bf 1408}, 051 (2014)
  [arXiv:1401.6088 [hep-th]].






\end{thebibliography}
\end{document}